\documentclass[aps,prb,preprint,showpacs]{revtex4}
\usepackage{amsmath}
\usepackage{amsfonts}
\usepackage{amssymb}%
\usepackage{graphicx,epsfig}
\begin{document}
\title{An ab-initio theory for the temperature dependence of magnetic anisotropy} 

\author{J.B.Staunton$^1$, L.Szunyogh $^{2,3}$, A.Buruzs$^{3}$,  B.L.Gyorffy $^{3,4}$, S.Ostanin$^{1}$ and
L.Udvardi $^{2,3}$ }

\affiliation{$^1$Department of Physics, University of Warwick, Coventry CV4 7AL, U.K.}

\affiliation{$^2$ Department of Theoretical Physics, Budapest 
University of Technology and Economics, Budapest, Hungary}

\affiliation{$^3$ Centre for Computational Material Science, Technical
University of Vienna, Getreidemarkt 9/134 A-1060, Vienna,  Austria}

\affiliation{$^4$ H.H.Wills Physics Laboratory, University of Bristol, Tyndall 
Avenue, Bristol BS8 1TL, U.K.}

\date{\today}

\begin{abstract}
We present a first-principles theory of the variation of 
magnetic anisotropy, $K$, with temperature, $T$, in metallic ferromagnets.
It is based on relativistic electronic structure theory and calculation of magnetic torque. 
Thermally induced `local moment' magnetic fluctuations are described within 
the relativistic generalisation of the `disordered local moment' (R-DLM) theory
from which the $T$ dependence of the magnetisation, $m$, is found. 
We apply the theory to a uniaxial magnetic material with {\it tetragonal} crystal
symmetry,$L1_0$-ordered FePd, and find its uniaxial $K$ consistent with a 
magnetic easy axis perpendicular to the Fe/Pd layers for all $m$ and 
proportional to $m^2$ for a broad range of values of $m$. This is the same
trend that we have previously found in $L1_0$-ordered FePt and which agrees with experiment.
This account, however, differs 
qualitatively from that extracted from a single ion anisotropy model.
 We also study the magnetically soft {\it cubic} magnet,
the $Fe_{50}Pt_{50}$ solid solution, and find that its small magnetic anisotropy constant $K_1$
rapidly diminishes from 8 $\mu$eV to zero. $K$ evolves from being proportional to $m^{7}$ at low $T$ 
to $m^4$ near the Curie temperature.

\end{abstract}
\maketitle

\section{INTRODUCTION}

 It is well-known that a description of magnetic anisotropy, K,
can be provided once relativistic effects such as the spin-orbit coupling
on the electronic structure of materials are considered. Over recent years `first-principles'
theoretical work, based on relativistic density functional theory, has been
quite successful in describing trends in K for a range of magnetic materials in
bulk, film and nanostructured form~\cite{Jansen,Kubler,review}, 
e.g.~\cite{Razee+99,Shick,Till,Laz,Qian,Cabria}. These results
can be fed into micromagnetic models of the magnetic properties to describe
phenomena such as magnetisation reversal processes in magnetic recording materials~\cite{micromag}. There are also implications for electronic transport effects such as anisotropic 
magnetoresistance (AMR)~\cite{AMR}.  Until only very recently, however,
the temperature dependence of K  was assumed to follow that given by single ion 
anisotropy models developed by Callen and Callen and others over 40 years ago~\cite{Callen}
This assumption has now been challenged by ab-initio electronic structure theory~\cite{Mryasov, MAEvsT}.
The unexpected dependence of the magnetic anisotropy of $L1_0$-FePt, found in experiment~\cite{Okamoto,Thiele2002,Wu}, to decrease in proportion with the square
of the magnetisation, $m(T)$, is described well by the new theoretical treatments whereas 
the single ion magnetic anisotropy models fail. Evidently the itinerant nature of the
electrons in metallic magnets like $FePt$ is an important factor. 

In this paper we present a detailed description of our ab-initio theory for the temperature dependence of magnetic anisotropy. It involves a fully relativistic description of the electronic structure and hence includes spin-orbit coupling effects. The thermally excited magnetic fluctuations are accounted for with the, by now, well-tried, 
disordered local moment (DLM) picture.~\cite{Moriya,DLM,JBS+BLG}

The study of temperature-dependent magnetic anisotropy has recently become particularly topical owing to extensive experimental
studies of magnetic films and nanostructures and 
their technological potential. For example, 
fabrication of assemblies of increasingly smaller magnetic
nanoparticles has great potential in the design of ultra-high
density magnetic data storage media.~\cite{Sun}  If thermally driven demagnetisation and loss of data is to be avoided
over a reasonable storage period, there is, however, 
a particle size limit to confront. A way of reducing this limit is to use 
materials with high magnetocrystalline anisotropy, $K$, since the 
superparamagnetic diameter of a magnetic particle is proportional to $(k_B T/K)
^{\frac{1}{3}}$, where $k_B T$ is the thermal energy.~\cite{OHandley}
Writing to media of very high $K$ material can be achieved by temporary 
heating.~\cite{Thiele2002,TAR} $K$
is reduced significantly during the magnetic write process and the 
information is locked in as the material cools. 
Modelling this process and improving the design of high density 
magnetic recording media therefore requires an understanding of how $K$ 
varies with temperature.

The temperature dependence of magnetic anisotropy 
in magnets where the magnetic moments are 
well-localised, e.g. rare-earth and oxide magnets, is described rather 
well by these single ion anisotropy models but it is
questionable whether this will also be the case for itinerant 
ferromagnets.~\cite{OHandley} Owing to its  high
uniaxial magnetocrystalline anisotropy (MCA) (4-10 10$^7$ ergs/cm$^3$ or up to 1.76 meV per $FePt$ pair~\cite{FePt-MAE,Farrow})
the chemically ordered $L1_0$ phase of equiatomic $FePt$, has 
attracted much attention as a potential ultra-high magnetic recording
density material. Indeed arrays of $FePt$ nanoparticles with
diameters as small as 3 nm have been synthesised.~\cite{Sun,Wu}
For a uniaxial magnet like this, $K$ is the difference between 
the free energies, 
$F^{(0,0,1)}$ and $F^{(1,0,0)}$ of the system magnetised along 
 $(0,0,1)$ and $(1,0,0)$ crystallographic directions. 
So for the first application of our theory we chose 
$L1_0$-ordered $FePt$.~\cite{MAEvsT}
Careful experimental studies of its fundamental magnetic 
properties.~\cite{Okamoto,Thiele2002,Wu} find that over a large temperature range, 
$K(T)/K(0)=(m(T)/m(0))^n$ ,where $n=2$ instead of $n=3$ as expected from
the simple single ion anisotropy model. We found our 
ab-initio calculations to be in good agreement with this 
surprising result. Mryasov et al.~\cite{Mryasov} independently examined
 the same issues with a different theoretical but complementary approach and drew the
same conclusions. In this paper, after providing full details of the R-DLM theory of magnetic anisotropy, we explore whether this $(m(T)/m(0))^2$
behavior is a general property of the MCA of $L1_0$-ordered itinerant
transition metal uniaxial magnets by investigating another 
important uniaxial magnetic material $FePd$. We also study the temperature dependence of 
a material which has cubic rather than the tetragonal crystal symmetry of $L1_0$-ordered
alloys, and which is magnetically softer, namely compositionally disordered 
$FePt$. 

In the next section we describe the temperature dependence of the magnetic
anisotropy that emerges from classical spin models with single site anisotropy. We then
review briefly current approaches to calculating $K$ from first-principles electronic theory
of materials at $T=0$K. An outline of the `disordered local moment' (DLM) picture of metallic
magnetism at finite temperature precedes a description of its relativistic generalisation. It is shown how the temperature dependence of the magnetisation, $m(T)$, can be found. The key outcome from the R-DLM theory is the formalism for the magnetisation dependence of magnetic anisotropy ab-initio. Applications to uniaxial $L1_0$-FePd
and cubic $Fe_{50}Pt_{50}$ follow and the final section provides a summary.

\section{SINGLE ION ANISOTROPY}
 
The MCA of a material can be conveniently expressed as
$K=\sum_{\gamma} K_{\gamma} g_{\gamma} ( \hat{n})$ where the $K_{\gamma}$'s are coefficients, $\hat{n}$ is the magnetisation
direction and $g_{\gamma}$'s are polynomials (spherical harmonics) of the 
angles $\vartheta$, $\varphi$, fixing the orientation of  
$\hat{n}$ with respect to the crystal axes, and belong to the fully symmetric representation of the 
crystal point group. 
As the temperature rises, $K$ decreases rapidly. The key features
of the results of the early theoretical work on this 
effect~\cite{Callen} are revealed by classical spin models pertinent to 
magnets with localised magnetic moments. The anisotropic behavior of 
a set of localised `spins' associated
with ions sitting on crystalline sites, $i$, in the material, is given by
a term in
the hamiltonian $H_{an} = \sum_i \sum_{\gamma} k_{\gamma} g_{\gamma} ( 
\widehat{e}_i)$ with 
$\widehat{e}_i$ a unit vector denoting the spin direction on the site $i$. 
As the
temperature is raised, the `spins' sample the energy surface over a small
angular range about the magnetisation direction and the anisotropy energy
is given from the difference between averages taken for the magnetisation 
along the easy and hard directions. If the coefficients $k_{\gamma}$ are assumed 
to be rather insensitive to temperature, the dominant thermal variation 
of $K$ for a ferromagnet is given by 
$K(T)/K(0) = \langle g_{l} ( \widehat{e})
\rangle_{T}/\langle g_{l} ( \widehat{e} ) \rangle_{0}$
The averages $\langle \cdots \rangle_{T}$ are taken such that $\langle 
\widehat{e}
\rangle_{T} = m(T)$, the magnetisation of the system at temperature $T$, and
 $l$ is the order of the spherical harmonic
describing the angular dependence of the local anisotropy i.e. $l =2$ and
$4$ for uniaxial and cubic systems respectively. At low temperatures 
$K(T)/K(0) \approx (m(T)/m(0))^{l(l+1)/2}$ and near the Curie 
temperature $T_c$, $K(T)/K(0) \approx (m(T)/m(0))^{l}$. 

These results can be illustrated straightforwardly in a way which will
be helpful for the development of our ab-initio theory.
 Consider a classical 
spin hamiltonian appropriate to a uniaxial ferromagnet.
\begin{equation}
H= -\frac{1}{2} \sum_{i,j} J_{ij} \widehat{e}_i \cdot \widehat{e}_j -k\sum_i 
(\widehat{n}_0 \cdot \widehat{e}_i)^2
\end{equation}
where $\widehat{e}_i$ describes the orientation of a classical spin at site
 $i$ and $J_{ij}$ and $k$ are exchange and anisotropy parameters. 
$\widehat{n}_0$ is a unit vector along the magnetic easy axis.
 A mean field description of the system is given by 
reference to a hamiltonian $\sum_i \vec{h} \cdot \widehat{e}_i$ 
where the orientation of Weiss field $\vec{h}$, i.e. 
$\vec{h}= h \widehat{n}$, determines the direction of the magnetisation 
of the system and has direction cosines ($\sin \vartheta \cos \varphi$,
$\sin \vartheta \sin \varphi$, $\cos \vartheta$). Within this mean 
field approximation the magnetisation $m$ is $
\vec{m}(T)= \int \widehat{e} \; P(\widehat{e})
d\widehat{e}$ where the
probability of a spin being orientated along $\widehat{e}$ is $P(\widehat{e})= 
e^{-\beta h \widehat{n} \cdot \widehat{e}}/Z_0$ with $Z_0= \int e^{-\beta h \widehat{n} \cdot \widehat{e}} d\widehat{e}$. 
The free energy 
difference per site  between the system magnetised along two directions 
$\widehat{n}_1$ and $\widehat{n}_2$ is
\begin{equation}
K(T) = -\frac{k}{Z_0} \int ((\widehat{n}_0 \cdot \widehat{e})^2 e^{-\beta h \widehat{n}_1
\cdot \widehat{e}} - (\widehat{n}_0\cdot \widehat{e})^2 e^{-\beta h 
\widehat{n}_2 
\cdot \widehat{e}}) d\widehat{e} 
\end{equation}
If $\widehat{n}_1$ and $\widehat{n}_2$ are parallel and 
perpendicular to the magnetic easy axis $\widehat{n}_0$ respectively then
\begin{equation}
K(T) = -\frac{k}{Z_0} \int g_2 (\widehat{n}_0 \cdot \widehat{e}) e^{-\beta h \widehat{n}_0 \cdot \widehat{e}} d\widehat{e}
\end{equation}
where $g_2$ is the Legendre polynomial $ (3(\widehat{n}_0 \cdot \widehat{e})^2 -1)/2$. As a function of the magnetisation $m(T)/m(0)$, $K(T)/K(0)$ varies
quadratically near the Curie temperature $T_c$ and cubically at low
$T$.  The same dependence can be shown for this simple spin model for 
the rate of variation of magnetic anisotropy with angle $\vartheta$ that the magnetisation makes with the system's easy axis, namely the magnetic 
torque~\cite{OHandley} $T_{\vartheta}= -\partial K/\partial \vartheta$.

\section{AB-INITIO THEORY OF MAGNETIC ANISOTROPY}
Magnetocrystalline anisotropy is caused largely by spin-orbit coupling and 
receives an ab-initio description from the relativistic generalisation of  
spin density functional (SDF) theory.~\cite{Jansen} 
Apart from the work by Mryasov et al.~\cite{Mryasov} and 
ourselves~\cite{MAEvsT}, up to now calculations of 
the anisotropy constants $K$ have been suited to $T=0$K only. 
Spin-orbit coupling effects are treated perturbatively
or with a fully relativistic theory~\cite{Razee+97,Razee+99}. 
Typically the total energy, 
or the single-electron contribution to it (if the force theorem
is used), is calculated for two or more magnetisation directions, 
$\hat{n}_1$ and $\hat{n}_2$ separately 
and then the MCA is obtained from the difference, $\Delta E$. $\Delta E$
is typically small ranging from meV to $\mu$eV and high precision in 
calculating the energies is required.
For example, we have used this rationale with a fully relativistic theory to study the 
MCA of magnetically soft, compositionally disordered binary and ternary 
component alloys~\cite{Razee+97,more} and the effect upon it of 
short-range~\cite{Razee+99} and long range chemical order~\cite{ICNDR}
in harder magnets such as $CoPt$ and $FePt$. 

Experimentally, measurements of magnetocrystalline anisotropy constants 
of magnets can be obtained from torque magnetometry~\cite{OHandley}. From similar
considerations of magnetic torque, ab-initio calculations of MCA can be 
made. There are obvious advantages in that the MCA can be obtained 
from a single calculation and reliance is not placed on the accurate 
extraction of 
a small difference between two energies. In particular the torque method
 has been used to good effect by Freeman and 
co-workers~\cite{Wang-torque} in conjunction with their state-tracking
method to study the MCA of a range of uniaxial magnets including 
layered systems.

If the free energy of a material magnetised along a direction specified
by $\widehat{n} =$ ( $\sin \vartheta \cos \varphi$,
$\sin \vartheta \sin \varphi $, $\cos \vartheta$ ) is 
$F^{(\widehat{n})}$, then the torque is
\begin{equation} 
\vec{T}^{(\widehat{n})} = - \frac{\partial F^{(\widehat{n})}}{\partial
\widehat{n}}.
\end{equation}
The contribution to the torque from the anisotropic part of 
$F^{(\widehat{n})}$ leads to a direct link
between the gap in the spin wave spectrum and the MCA by the solution 
of the equation~\cite{Russian-spinwave} 
\begin{equation}
\frac{d \widehat{n}}{dt} = \gamma (\widehat{n} \wedge 
\vec{T}^{(\widehat{n})}).
\end{equation}
where $\gamma$ is the gyromagnetic ratio.
Closely related to $\vec{T}^{(\widehat{n})}$ is the variation of 
$F^{(\widehat{n})}$ with respect to $\vartheta$ and $\varphi$, i.e.
$T_{\vartheta}(\vartheta,\varphi)= -\frac{\partial F^{(\widehat{n})}}
{\partial \vartheta}$ and $T_{\varphi}(\vartheta,\varphi)= 
-\frac{\partial F^{(\widehat{n})}}{\partial \varphi}$. As shown 
by Wang et al.~\cite{Wang-torque}, for most uniaxial magnets, which are
 well approximated by a free energy of the form
\begin{equation}
F^{(\widehat{n})} = F_{iso} + K_2 \sin^2 \vartheta +
K_4 \sin^4 \vartheta , 
\end{equation}
(where $K_2$ and $K_4$ and magnetocrystalline
anisotropy constants and $F_{iso}$ is the isotropic part of the free
energy), $T_{\vartheta}(\vartheta= \pi/4, \varphi =0) = -
(K_2 +K_4)$. This is equal to the MCA, $\Delta F = 
F^{(1,0,0)}- F^{(0,0,1)}$. For a magnet with cubic symmetry so that
\begin{equation}
F^{(\widehat{n})} \approx F_{iso} + K_1 ( \sin^4 \vartheta \sin^2 2 
\varphi + \sin^2 2 \vartheta),\label{MAE-cubic}
\end{equation}
a calculation of $T_{\varphi} ( 
\vartheta= \pi/2, \varphi= \pi/8)$ gives $-K_1/2$. In this work we 
present our formalism for the direct calculation of the torque 
quantities $T_{\vartheta} (\vartheta, \varphi)$ and 
$T_{\varphi}(\vartheta, \varphi)$, and hence
the MCA, in which the effects of thermally induced magnetic fluctuations
are included so that the temperature dependence is captured.

In our formalism the motion of the electron is described with
 spin-polarised, relativistic multiple scattering theory. An adaptive 
mesh algorithm~\cite{EB+BG} for Brillouin 
zone integrations is used in the calculations to ensure adequate
 numerical precision for the MCA to within 
0.1~$\mu$eV~\cite{Razee+97,Razee+99}. Since we characterise the 
thermally induced magnetic fluctuations in terms of disordered local 
moments, we now go on to describe this picture of finite temperature 
magnetism.

\section{METALLIC MAGNETISM AT FINITE TEMPERATURES - `DISORDERED LOCAL 
MOMENTS}

In a metallic ferromagnet at $T=0$K the electronic band structure is
spin-polarised. With increasing temperature, spin fluctuations are induced 
which eventually
destroy the long-range magnetic order and hence the overall spin polarization 
of the system's electronic structure. These collective electron modes interact
as the temperature is raised and are dependent upon and affect the underlying
electronic structure. For many materials the magnetic excitations can be
modelled by associating local spin-polarisation axes with all lattice sites and
the orientations $\{ \hat{e} \}$ vary very slowly on the time-scale of the
electronic motions.~\cite{Moriya} These 
`local moment' degrees of freedom produce local 
magnetic fields on the lattice sites which affect the electronic motions and
are self-consistently maintained by them. By taking appropriate ensemble 
averages over the orientational configurations, the system's 
magnetic properties can be determined.   

This `disordered local moment' (DLM) picture has been implemented within
a multiple-scattering (Korringa-Kohn-Rostoker, KKR)~\cite{KKR,KKR-CPA,scf-kkr-cpa}
 formalism using the
first-principles approach to the problem of itinerant electron magnetism.
 At no stage does it map the many-electron problem onto an effective 
Heisenberg model, and yet it deals, quantitatively, with both the 
ground state and the demise of magnetic long-range order at the Curie 
temperature in a material-specific, parameter-free manner. It 
has been used to describe the experimentally observed local exchange 
splitting and magnetic short-range order in both ultra-thin Fe films~\cite{fccFe-2002}
 and bulk Fe, the damped RKKY-like magnetic interactions in the 
compositionally disordered CuMn 'spin-glass' alloys~\cite{MFL-EPL} and 
the quantitative description of the onset of magnetic order in a 
range of alloys~\cite{DLM-alloys,invar-Entel}. In combination with the 
local self-interaction correction (L-SIC)~\cite{LSIC} for strong 
electron correlation effects, it also gives a revealing account of 
 magnetic ordering in rare earths~\cite{Ian-Gd}.  Other applications of
 the DLM picture include dilute magnetic semiconductors~\cite{Dederichs}
and actinides~\cite{sweden}. Short-range order of the local moments can
be explicitly included by making use of the recently developed KKR-
nonlocal-CPA (KKR-NLCPA)~\cite{Rowlands,NLCPA-totE}.

We now briefly recap on how this `disordered local moment' (DLM) picture
is implemented using the KKR-CPA and how a ferromagnetic metal both 
above and below the Curie temperature can be described. Our main objective in this paper is to explain its relativistic extension 
and show how this leads to an ab-initio theory of the temperature 
dependence of magnetic anisotropy when relativistic effects are 
explicitly included.

\section{RELATIVISTIC DISORDERED LOCAL MOMENT THEORY}

\subsection{\bigskip General framework}
The non-relativistic version of the DLM theory has been discussed in 
detail by Gyorffy et al.~\cite{DLM,JBS+BLG} Here we summarise 
the general framework and concentrate on those aspects which are 
necessary for a
description of magnetic anisotropy.  The starting point is the 
specification of $ \Omega^{(\widehat{n})} ( \{ \widehat{e} \} ) $, the 
`generalised' electronic grand potential taken from the relativistic
extension of spin density functional theory (SDFT)~\cite{DLM,Jansen}. It specifies
an itinerant electron system constrained such that the local spin
polarisation axes are configured according to 
$\{\widehat{e} \}=\{\widehat{e}_{1},\widehat{e}_{2},\ldots,\widehat{e}_{N}\}$ where $N$ is the number of sites (moments) in the system. 
For magnetic anisotropy to be described, relativistic effects
such as spin-orbit coupling upon the motion of the electrons must be
included. This means that orientations of the local moments with 
respect to a specified direction $\hat{n}$ within the material
are relevant. The role of a (classical) local moment
 hamiltonian, albeit a highly complicated one, is played by 
$\Omega^{(\widehat{n})} (\{ \widehat{e} \} )$. Note that in the 
following we do not prejudge the physics by trying to extract an 
effective `spin' model from $\Omega^{(\widehat{n})} (\{ 
\widehat{e} \} )$ such as a 
classical Heisenberg model with a single site anisotropy term.   

Consider a ferromagnetic metal magnetised along a
direction $\widehat{n}$ at a temperature $T$
where the orientational probability distribution is denoted by
$P^{\left(  \widehat{n}\right)  }\left(  \left\{  \widehat{e}\right\}
\right)  ,$ and its average 
\begin{equation}
\left\langle \widehat{e}_{i}\right\rangle =\int\!\dots\!\int\widehat{e}%
_{i}P^{\left(  \widehat{n}\right)  }\left(  \left\{  \widehat{e}\right\}
\right)  \,d\widehat{e}_{1}\dots d\widehat{e}_{N}=\widehat{n}\;.
\end{equation}
is aligned with the magnetisation direction $\widehat{n}$.
The
canonical partition function and the probability function are defined as%
\begin{equation}
Z^{\left(  \widehat{n}\right)  }=\int\!\dots\!\int e^{-\beta \Omega^{\left(
\widehat{n}\right)  }\left(  \left\{  \widehat{e}\right\}  \right)
}\,d\widehat{e}_{1}\dots d\widehat{e}_{N}\;,
\end{equation}
and
\begin{equation}
P^{\left(  \widehat{n}\right)  }\left(  \left\{  \widehat{e}\right\}  \right)
=\frac{e^{-\beta \Omega^{\left(  \widehat{n}\right)  }\left(  \left\{  \widehat
{e}\right\}  \right)  }}{Z^{\left(  \widehat{n}\right)  }}\quad,
\end{equation}
respectively. The thermodynamic free-energy which 
includes the entropy associated with the orientational fluctuations as 
well as creation of electron-hole pairs, is given by%
\begin{equation}
F^{\left(  \widehat{n}\right)  }=-\frac{1}{\beta}\ln Z^{\left(  \widehat
{n}\right)  }\;.
\end{equation}
By choosing a trial Hamiltonian function, $\Omega_{0}^{\left(
\widehat{n}\right)  }\left(  \left\{  \widehat{e}\right\}  \right)  $ 
with $
Z_{0}^{\left(  \widehat{n}\right)  }=\int\!\dots\!\int e^{-\beta
\Omega_{0}^{\left(  \widehat{n}\right)  }\left(  \left\{  \widehat{e}\right\}
\right)  }\,d\widehat{e}_{1}\dots d\widehat{e}_{N}$, 
\begin{equation}
P_{0}^{\left(  \widehat{n}\right)  }\left(  \left\{  \widehat{e}\right\}
\right)  =\frac{e^{-\beta \Omega_{0}^{\left(  \widehat{n}\right)  }\left(  \left\{
\widehat{e}\right\}  \right)  }}{Z_{0}^{\left(  \widehat{n}\right)  }}%
\end{equation}
and $
F_{0}^{\left(  \widehat{n}\right)  }=-\frac{1}{\beta}\ln Z_{0}^{\left(
\widehat{n}\right)  }$ 
the \emph{Feynman-Peierls Inequality}~\cite{Feynman} implies an upper
bound for the free energy, i.e.,
\begin{equation}
F^{\left(  \widehat{n}\right)  }\leq F_{0}^{\left(  \widehat{n}\right)
}+\left\langle \Omega^{\left(  \widehat{n}\right)  }-\Omega_{0}^{\left(  \widehat
{n}\right)  }\right\rangle ^{0}\;,
\end{equation}
where the average refers to the probability $P_{0}^{\left(  \widehat
{n}\right)  }\left(  \left\{  \widehat{e}\right\}  \right)  $. By expanding
$\Omega_{0}^{\left(  \widehat{n}\right)  }\left(  \left\{  \widehat{e}\right\}
\right)  $ as%
\begin{equation}
\Omega_{0}^{\left(  \widehat{n}\right)  }\left(  \left\{  \widehat{e}\right\}
\right)  =\sum_{i}\omega_{i}^{1\left(  \widehat{n}\right)  }\left(  \widehat{e}%
_{i}\right)  +\frac{1}{2}\sum_{i\neq i}\omega_{i,j}^{2\left(  \widehat{n}\right)
}\left(  \widehat{e}_{i},\widehat{e}_{j}\right)  +\ldots\;,
\label{H-expansion}%
\end{equation}
the `best' trial system is found to satisfy~\cite{DLM,JBS+BLG}%
\begin{equation}
\left\langle \Omega^{\left(  \widehat{n}\right)  }\right\rangle _{\widehat{e}_{i}%
}^{0}-\left\langle \Omega^{\left(  \widehat{n}\right)  }\right\rangle
^{0}=\left\langle \Omega_{0}^{\left(  \widehat{n}\right)  }\right\rangle
_{\widehat{e}_{i}}^{0}-\left\langle \Omega_{0}^{\left(  \widehat{n}\right)
}\right\rangle ^{0}\;, \label{condition-1}%
\end{equation}%
\begin{equation}
\left\langle \Omega^{\left(  \widehat{n}\right)  }\right\rangle _{\widehat{e}%
_{i},\widehat{e}_{j}}^{0}-\left\langle \Omega^{\left(  \widehat{n}\right)
}\right\rangle ^{0}=\left\langle \Omega_{0}^{\left(  \widehat{n}\right)
}\right\rangle _{\widehat{e}_{i},\widehat{e}_{j}}^{0}-\left\langle
\Omega_{0}^{\left(  \widehat{n}\right)  }\right\rangle ^{0}\;, \label{condition-2}%
\end{equation}
and so on, where $\left\langle \quad\right\rangle _{\widehat{e}_{i}}$ or
$\left\langle \quad\right\rangle _{\widehat{e}_{i},\widehat{e}_{j}}$ denote
restricted statistical averages with $\widehat{e}_{i}$ or both 
$\widehat{e}_{i}$ and $\widehat{e}_{j}$ kept fixed, respectively. For example,
for a given physical quantity, $X$,
\begin{equation}
\left\langle X^{\left(  \widehat{n}\right)  }\right\rangle _{\widehat{e}_{i}%
}=\frac{\int\!\dots\!\int X^{\left(  \widehat{n}\right)  }\left(  \left\{
\widehat{e}\right\}  \right)  P_{0}^{\left(  \widehat{n}\right)  }\left(
\left\{  \widehat{e}\right\}  \right)  \,d\widehat{e}_{1}\dots d\widehat
{e}_{i-1}d\widehat{e}_{i+1}\dots d\widehat{e}_{N}\;}{P_{i}^{\left(  \widehat
{n}\right)  }\left(  \widehat{e}_{i}\right)  },
\end{equation}
with%
\begin{equation}
P_{i}^{\left(  \widehat{n}\right)  }\left(  \widehat{e}_{i}\right)
=\int\!\dots\!\int P_{0}^{\left(  \widehat{n}\right)  }\left(  \left\{
\widehat{e}\right\}  \right)  \,d\widehat{e}_{1}\dots d\widehat{e}%
_{i-1}d\widehat{e}_{i+1}\dots d\widehat{e}_{N}\;.
\end{equation}
The relationship,%
\begin{equation}
\left\langle X^{\left(  \widehat{n}\right)  }\right\rangle =\int\left\langle
X^{\left(  \widehat{n}\right)  }\right\rangle _{\widehat{e}_{i}}%
\,P_{i}^{\left(  \widehat{n}\right)  }\left(  \widehat{e}_{i}\right)
d\widehat{e}_{i}\;,
\end{equation}
is then obviously satisfied.

\subsection{Mean-field theory}

In this case we take a trial system which is comprised of the first 
term in Eq. (\ref{H-expansion}) only,%
\begin{equation}
\Omega_{0}^{\left(  \widehat{n}\right)  }\left(  \left\{  \widehat{e}\right\}
\right)  =\sum_{i}\omega_{i}^{\left(  \widehat{n}\right)  }\left(  \widehat{e}%
_{i}\right)  \;.
\end{equation}
Therefore, the quantities defined above reduce to%
\begin{equation}
Z_{0}^{\left(  \widehat{n}\right)  }=\int\!\dots\!\int%
{\displaystyle\prod\limits_{i}}
e^{-\beta \omega_{i}^{\left(  \widehat{n}\right)  }\left(  \widehat{e}_{i}\right)
}\,d\widehat{e}_{1}\dots d\widehat{e}_{N}\;=%
{\displaystyle\prod\limits_{i}}
Z_{i}^{\left(  \widehat{n}\right)  }\;,\;Z_{i}^{\left(  \widehat{n}\right)
}=\int e^{-\beta \omega_{i}^{\left(  \widehat{n}\right)  }\left(  \widehat{e}%
_{i}\right)  }\,d\widehat{e}_{i}\;,
\end{equation}%
\begin{equation}
P_{0}^{\left(  \widehat{n}\right)  }\left(  \left\{  \widehat{e}\right\}
\right)  =%
{\displaystyle\prod\limits_{i}}
P_{i}^{\left(  \widehat{n}\right)  }\left(  \widehat{e}_{i}\right)
\;,\;P_{i}^{\left(  \widehat{n}\right)  }\left(  \widehat{e}_{i}\right)
=\frac{e^{-\beta \omega_{i}^{\left(  \widehat{n}\right)  }\left(  \widehat{e}%
_{i}\right)  }}{Z_{i}^{\left(  \widehat{n}\right)  }}\;,
\end{equation}
and%
\begin{equation}
F_{0}^{\left(  \widehat{n}\right)  }=-\frac{1}{\beta}\sum_{i}\ln
Z_{i}^{\left(  \widehat{n}\right)  }\;.
\end{equation}
In order to employ condition (\ref{condition-1}) the following averages have
to evaluated,%
\begin{equation}
\left\langle \Omega_{0}^{\left(  \widehat{n}\right)  }\right\rangle ^{0}=\sum_{i}%
{\displaystyle\prod\limits_{j}}
\int\! \omega_{i}^{\left(  \widehat{n}\right)  }\left(  \widehat{e}_{i}\right)
P_{j}^{\left(  \widehat{n}\right)  }\left(  \widehat{e}_{j}\right)
d\widehat{e}_{j}=\sum_{i}\int\! \omega_{i}^{\left(  \widehat{n}\right)  }\left(
\widehat{e}_{i}\right)  P_{i}^{\left(  \widehat{n}\right)  }\left(
\widehat{e}_{i}\right)  d\widehat{e}_{i}\;,
\end{equation}%
\begin{equation}
\left\langle \Omega_{0}^{\left(  \widehat{n}\right)  }\right\rangle _{\widehat
{e}_{i}}^{0}=\omega_{i}^{\left(  \widehat{n}\right)  }\left(  \widehat{e}%
_{i}\right)  +\sum_{j\neq i}\int\!\omega_{j}^{\left(  \widehat{n}\right)  }\left(
\widehat{e}_{j}\right)  P_{j}^{\left(  \widehat{n}\right)  }\left(
\widehat{e}_{j}\right)  d\widehat{e}_{j}\;,
\end{equation}
consequently, $
\left\langle \Omega_{0}^{\left(  \widehat{n}\right)  }\right\rangle _{\widehat
{e}_{i}}^{0}-\left\langle \Omega_{0}^{\left(  \widehat{n}\right)  }\right\rangle
^{0}=\omega_{i}^{\left(  \widehat{n}\right)  }\left(  \widehat{e}_{i}\right)
-\int\!\omega_{i}^{\left(  \widehat{n}\right)  }\left(  \widehat{e}_{i}^{\prime
}\right)  P_{i}^{\left(  \widehat{n}\right)  }\left(  \widehat{e}_{i}^{\prime
}\right)  d\widehat{e}_{i}^{\prime}$ 
 and Eq. (\ref{condition-1}) implies,%
\begin{equation}
\omega_{i}^{\left(  \widehat{n}\right)  }\left(  \widehat{e}_{i}\right)
-\left\langle \Omega^{\left(  \widehat{n}\right)  }\right\rangle _{\widehat{e}_{i}%
}^{0}=\int\left(  \!\omega_{i}^{\left(  \widehat{n}\right)  }\left(  \widehat
{e}_{i}^{\prime}\right)  -\left\langle \Omega^{\left(  \widehat{n}\right)
}\right\rangle _{\widehat{e}_{i}^{\prime}}^{0}\right)  P_{i}^{\left(
\widehat{n}\right)  }\left(  \widehat{e}_{i}^{\prime}\right)  d\widehat{e}%
_{i}^{\prime}=const.\;,
\end{equation}
which leads to the relationship%
\begin{equation}
\omega_{i}^{\left(  \widehat{n}\right)  }\left(  \widehat{e}_{i}\right)
=\left\langle \Omega^{\left(  \widehat{n}\right)  }\right\rangle _{\widehat{e}_{i}%
}^{0}\;.
\end{equation}
The free energy is now given by 
\begin{equation}
F^{\left(  \widehat{n}\right)  }= 
\left\langle \Omega^{\left(  \widehat{n}\right)  }\right\rangle^0 + 
\frac{1}{\beta} \sum_i \int P_{i}^{\left(  \widehat{n}\right) 
 }\left(  \widehat{e}_{i}\right)\ln P_{i}^{\left(  
\widehat{n}\right)  }\left(  \widehat{e}_{i}\right) 
d\widehat{e}_{i} \;. \label{FreeEnergy}%
\end{equation}
This is the key expression for our subsequent development of the 
magnetic anisotropy energy.
(In the following we shall omit the superscript $0$ from the averages.)

\subsection{The Weiss field}

To proceed further one can expand $\omega_{i}^{\left(  \widehat{n}\right)  }\left(
\widehat{e}\right)  $ in terms of spherical harmonics,%
\begin{equation}
\omega_{i}^{\left(  \widehat{n}\right)  }\left(  \widehat{e}\right)  =\sum_{\ell
,m}\omega_{i,\ell m}^{\left(  \widehat{n}\right)  }\,Y_{\ell m}\left(  \widehat
{e}\right)  \;,
\end{equation}
where the constant term, $\frac{1}{\sqrt{4\pi}}\omega_{i,00}^{\left(  \widehat
{n}\right)  }$, does not enter the statistical averages and therefore
 can be
taken to be zero. The coefficients, $\omega_{i,\ell m}^{\left(  \widehat{n}\right)
}\left(  \widehat{e}_{i}\right)  $, can obviously be expressed as%
\begin{equation}
\omega_{i,\ell m}^{\left(  \widehat{n}\right)  }=\int \omega_{i}^{\left(  \widehat
{n}\right)  }\left(  \widehat{e}_{i}\right)  \,Y_{\ell m}^{\ast}\left(
\widehat{e}_{i}\right)  d\widehat{e}_{i}=\int\left\langle \Omega^{\left(
\widehat{n}\right)  }\right\rangle _{\widehat{e}_{i}}\,Y_{\ell m}^{\ast
}\left(  \widehat{e}_{i}\right)  d\widehat{e}_{i}\;.
\end{equation}

Keeping the leading term only, $\ell=1$, with $
Y_{1,\pm1}\left(  \widehat{e}_{i}\right)  =\sqrt{\frac{3}{8\pi}}\left(
e_{x}\pm ie_{y}\right)  \;,\;Y_{1,\pm1}\left(  \widehat{e}_{i}\right)
=\sqrt{\frac{3}{4\pi}}e_{z}$, 
$\omega_{i}^{\left(  \widehat{n}\right)  }\left(  \widehat{e}_{i}\right)  $ can be written as $
\omega_{i}^{\left(  \widehat{n}\right)  }\left(  \widehat{e}_{i}\right)  =\vec
{h}_{i}^{\left(  \widehat{n}\right)  }\cdot\widehat{e}_{i}
$
with%
\begin{equation}
\vec{h}_{i}^{\left(  \widehat{n}\right)  }=\int\frac{3}{4\pi}\widehat{e}%
_{i}\left\langle \Omega^{\left(  \widehat{n}\right)  }\right\rangle _{\widehat
{e}_{i}}\,d\widehat{e}_{i}\;.
\end{equation}

Furthermore,%
\begin{align}
Z_{i}^{\left(  \widehat{n}\right)  }  &  =\int\exp\left(  -\beta\vec{h}_{i}^{\left(  \widehat
{n}\right)  }\cdot\widehat{e}_{i}\right)  d\widehat{e}_{i}\\
&  = \frac{4\pi}{\beta
h_{i}^{\left(  \widehat{n}\right)  }}\sinh\beta h_{i}^{\left(  \widehat
{n}\right)  }\;,
\end{align}
where $
h_{i}^{\left(  \widehat{n}\right)  }=\left\vert \vec{h}_{i}^{\left(
\widehat{n}\right)  }\right\vert $, and the probability distribution is

\begin{equation}
P_{i}^{\left(  \widehat{n}\right)  }\left(  \widehat{e}_{i}\right)
=\frac{\beta h_{i}^{\left(  \widehat{n}\right)  }}{4\pi\sinh\beta
h_{i}^{\left(  \widehat{n}\right)  }}\exp\left(  -\beta\vec{h}_{i}^{\left(
\widehat{n}\right)  }\cdot\widehat{e}_{i}\right)  \;.
\end{equation}
Thus the average alignment of the local moments, proportional to the 
magnetisation, is
\begin{align}
\vec{m}_{i}^{\left(  \widehat{n}\right)  }  &  =\left\langle \widehat{e}%
_{i}\right\rangle =\frac{\beta h_{i}^{\left(  \widehat{n}\right)  }}{4\pi
\sinh\beta h_{i}^{\left(  \widehat{n}\right)  }}\int\widehat{e}_{i}\exp\left(
-\beta\vec{h}_{i}^{\left(  \widehat{n}\right)  }\cdot\widehat{e}_{i}\right)
d\widehat{e}_{i}. \label{mag}
\end{align}
from which
$\vec{m}_{i}^{\left(  \widehat{n}\right)  }=m_{i}^{\left(  \widehat{n}\right)
}\,\widehat{h}_{i}^{\left(  \widehat{n}\right)  }$
and
\begin{equation}
m_{i}^{\left(  \widehat{n}\right)  }=-\frac{d\ln Z_{i}^{\left(  \widehat
{n}\right)  }}{d\left(  \beta h_{i}^{\left(  \widehat{n}\right)  }\right)
}=\frac{1}{\beta h_{i}^{\left(  \widehat{n}\right)  }}-\coth\beta
h_{i}^{\left(  \widehat{n}\right)  }=L\left(  -\beta h_{i}^{\left(
\widehat{n}\right)  }\right)  \label{moment}%
\end{equation}
follow, where $L(x)$ is the Langevin function. Since in the ferromagnetic
state, $
\widehat{m}_{i}=\frac{\vec{m}_{i}}{m_{i}}=\widehat{n}$
we finally can write the Weiss field,$
\vec{h}_{i}^{\left(  \widehat{n}\right)  }=h_{i}^{\left(  
\widehat{n}\right)}\,\widehat{n}$ as
\begin{equation}
h_{i}^{\left(  \widehat{n}\right)  }=\frac{3}{4\pi}\int\left(  \widehat{e}%
_{i}\cdot\widehat{n}\right)  \left\langle \Omega^{\left(  \widehat{n}\right)
}\right\rangle _{\widehat{e}_{i}}\,d\widehat{e}_{i}\;. \label{WF-1}%
\end{equation}

Note that an identical Weiss field $\vec{h}^{\left(  \widehat{n}\right)}$ 
associated with every site corresponds to a
description of a ferromagnetic system magnetised along $\widehat{n}$ 
with no reference to an external field.

\subsection{Averaging With The Coherent Potential Approximation}

In order to calculate the restricted average, $\left\langle \Omega^{\left(
\widehat{n}\right)  }\right\rangle _{\widehat{e}_{i}}$, from first principles, as discussed by Gyorffy et al.~\cite{DLM} 
we follow the strategy of the Coherent Potential Approximation (CPA)~\cite{cpa} as
combined with the KKR method~\cite{KKR-CPA}. The electronic charge 
density and also the 
magnetisation density, which sets the magnitudes, $\{ \mu \}$, of the
local moments, are determined from a self-consistent field
(SCF)-KKR-CPA~\cite{scf-kkr-cpa} 
calculation. For the systems we discuss in this paper, the magnitudes
 of the local moments are rather insensitive to the orientations of the
local moments surrounding them~\cite{JBS+BLG}. We return to this point
later. 

For a given set of (self-consistent) potentials, electronic charge and
local moment magnitudes $\{ \mu_i \}$, the orientations of the local 
moments are accounted for by the similarity transformation of the 
single-site t-matrices~\cite{Messiah},
\begin{equation}
\underline{t}_{i} \left(  \widehat{e}%
_{i}\right)  =\underline{R}\left(  \widehat{e}_{i}\right)  \,\underline{t}%
_{i} \left(  \widehat{z}\right)  \underline
{R}\left(  \widehat{e}_{i}\right)  ^{+}\;,
\end{equation}
where for a given energy (not labelled explicitly) $\underline{t}_{i} \left(  \widehat{z}\right)  $ stands for the
\emph{t}-matrix with effective field pointing along the local $z$ axis~\cite{Strange1984} and
$\underline{R}\left(  \widehat{e}_{i}\right)  $ is a unitary representation of
the $O\left(  3\right)  $ transformation that rotates the $z$ axis 
along $\widehat{e}_{i}$. In this work $\underline{t}_{i} \left(  
\widehat{z}\right)  $ is found by considering the relativistic,
spin-polarised scattering of an electron from a central potential with
a magnetic field defining the z-axis~\cite{Strange1984}. Thus 
spin-orbit coupling effects are naturally included. 

The \emph{single-site CPA} determines an effective medium through which the
motion of an electron mimics the motion of an electron \emph{on the average}. In a system
magnetised along a direction $\widehat{n}$, the medium is specified by 
 \emph{t}-matrices, $\underline
{t}_{i,c}^{\left(  \widehat{n}\right)  }$, which satisfy the condition
~\cite{KKR-CPA},
\begin{equation}
\left\langle \underline{\tau}_{ii}^{\left(  \widehat{n}\right)  }\left(
\left\{  \widehat{e}\right\}  \right)  \right\rangle =\int\left\langle
\underline{\tau}_{ii}^{\left(  \widehat{n}\right)  }\right\rangle
_{\widehat{e}_{i}}P_{i}^{\left(  \widehat{n}\right)  }\left(  \widehat{e}%
_{i}\right)  d\widehat{e}_{i}=\underline{\tau}_{ii,c}^{\left(  \widehat
{n}\right)  }\;, \label{CPA-1}%
\end{equation}
where the site-diagonal matrices of the multiple scattering path 
operator~\cite{Gyorffy+Stott} are defined as,%
\begin{equation}
\left\langle \underline{\tau}_{ii}^{\left(  \widehat{n}\right)  }\right\rangle
_{\widehat{e}_{i}}=\underline{\tau}_{ii,c}^{\left(  \widehat{n}\right)
}\underline{D}_{i}^{\left(  \widehat{n}\right)  }\left(  \widehat{e}%
_{i}\right)  \;,
\end{equation}%
\begin{equation}
\underline{D}_{i}^{\left(  \widehat{n}\right)  }\left(  \widehat{e}%
_{i}\right)  =\left(  \underline{1}+\left[  \left(  \underline{t}_{i}
\left(  \widehat{e}_{i}\right)  \right)  ^{-1}-\left(
\underline{t}_{i,c}^{\left(  \widehat{n}\right)  }\right)  ^{-1}\right]
\underline{\tau}_{ii,c}^{\left(  \widehat{n}\right)  }\right)  ^{-1}\;,
\label{Dmatrix}%
\end{equation}
and%
\begin{equation}
\underline{\underline{\tau}}_{c}^{\left(  \widehat{n}\right)  }=\left(
\left(  \underline{\underline{t}}_{c}^{\left(  \widehat{n}\right)  }\right)
^{-1}-\underline{\underline{G}}_{0}\right)  ^{-1}\;. \label{tau-CPA}%
\end{equation}
In the above equation, double underlines denote matrices in site
and angular
momentum space. $\underline{\underline{t}}_{c}^{\left(  \widehat{n}\right)  }$
is diagonal with respect to site indices, while $\underline{\underline{G}}_{0}$ stands for
the matrix of structure constants~\cite{KKR, EB+BG}.
Eq. (\ref{CPA-1}) can be  rewritten 
by introducing the excess scattering matrices,
\begin{equation}
\underline{X}_{i}^{\left(  \widehat{n}\right)  }\left(  \widehat{e}%
_{i}\right)  =\left(  \left[  \left(  \underline{t}_{i,c}^{\left(  \widehat
{n}\right)  }\right)  ^{-1}-\left(  \underline{t}_{i}\left(  \widehat{e}_{i}\right)  \right)  ^{-1}\right]
^{-1}-\underline{\tau}_{ii,c}^{\left(  \widehat{n}\right)  }\right)
^{-1}\;,\label{Xmatrix}%
\end{equation}
in the form%
\begin{equation}
\int\underline{X}_{i}^{\left(  \widehat{n}\right)  }\left(  \widehat{e}%
_{i}\right)  P_{i}^{\left(  \widehat{n}\right)  }\left(  \widehat{e}%
_{i}\right)  d\widehat{e}_{i}=\underline{0}\;.\label{CPA-3}%
\end{equation}

Thus, for a given set of Weiss fields, $h_{i}^{\left(  \widehat{n}\right)  },$
and corresponding probabilities,%
\begin{equation}
P_{i}^{\left(  \widehat{n}\right)  }\left(  \widehat{e}_{i}\right)
=\frac{\beta h_{i}^{\left(  \widehat{n}\right)  }}{4\pi\sinh\beta
h_{i}^{\left(  \widehat{n}\right)  }}\exp\left(  -\beta h_{i}^{\left(
\widehat{n}\right)  }\widehat{n}\cdot\widehat{e}_{i}\right)  \;, \label{Pi}%
\end{equation}
Eq. (\ref{CPA-3}) can be solved by iterating together with Eqs. (\ref{Xmatrix}%
) and (\ref{tau-CPA}) to obtain the matrices, $\underline{t}_{i,c}^{\left(
\widehat{n}\right)  }$. 
The integral in Eq. (\ref{CPA-3}) can be discretized to yield a multi-component CPA equation
which can be solved by the method proposed by Ginatempo and 
Staunton~\cite{Ginatempo+JBS}.
Care has to be taken, in particular for low temperatures where 
$P_{i}^{\left(  \widehat{n}\right)  }\left(  \widehat{e}%
_{i}\right)$ is a sharply structured function, to include a large number 
 and/or an adaptive sampling of the grid points.

\subsection{Calculation of the Weiss field}

In the spirit of the \emph{magnetic force theorem} we shall consider only the
single--particle energy (band energy) part of the SDFT Grand Potential 
as an effective `local moment' Hamiltonian in Eq. (\ref{WF-1}),
\begin{align}
\Omega^{\left(  \widehat{n}\right)  }\left(  \left\{  \widehat{e}\right\}
\right)   & \simeq-\int d\varepsilon\,f_{FD}\left(  
\varepsilon;\nu^{\left(
\widehat{n}\right)  }\right)  \,N^{\left(  \widehat{n}\right)  }\left(
\varepsilon;\left\{  \widehat{e}\right\}  \right)  \;,
\end{align}
where $\nu^{\left(  \widehat{n}\right)  }$ is the chemical potential,
$f_{FD}\left(  \varepsilon;\nu^{\left(  \widehat{n}\right)  }\right)  $ is the
Fermi-Dirac distribution, and $N^{\left(  \widehat{n}\right)  }\left(
\varepsilon;\left\{  \widehat{e}\right\}  \right)  $ denotes the integrated
density of states for the orientational configuration, $\left\{  \widehat
{e}\right\}  $. For a `good local moment' system such as many iron and
cobalt alloys, this frozen potential approximation is well-justified
and discussed shortly.

The Lloyd formula~\cite{Lloyd} provides an explicit expression for
$N^{\left(  \widehat{n}\right)  }\left(  \varepsilon;\left\{  \widehat
{e}\right\}  \right)  ,$%
\begin{equation}
N^{\left(  \widehat{n}\right)  }\left(  \varepsilon;\left\{  \widehat
{e}\right\}  \right)  =N_{0}\left(  \varepsilon\right)  -\frac{1}{\pi
}\operatorname{Im}\ln\det\left(  \underline{\underline{t}}^{\left(
\widehat{n}\right)  }\left(  \varepsilon;\left\{  \widehat{e}\right\}
\right)  ^{-1}-\underline{\underline{G}}_{0}\left(  \varepsilon\right)
\right)  \,,
\end{equation}
with $N_{0}\left(  \varepsilon\right)  $ being the integrated DOS of the free
particles. The integrated DOS can be further decomposed into two more
pieces%
\begin{equation}
N^{\left(  \widehat{n}\right)  }\left(  \varepsilon;\left\{  \widehat
{e}\right\}  \right)  =N_{0}\left(  \varepsilon\right)  +N_{c}^{\left(
\widehat{n}\right)  }\left(  \varepsilon\right)  +\Delta N^{\left(
\widehat{n}\right)  }\left(  \varepsilon;\left\{  \widehat{e}\right\}
\right)  \;,
\end{equation}
where%
\begin{equation}
N_{c}^{\left(  \widehat{n}\right)  }\left(  \varepsilon\right)  =\frac{1}{\pi
}\operatorname{Im}\ln\det\underline{\underline{\tau}}_{c}^{\left(  \widehat
{n}\right)  }\left(  \varepsilon\right)  =-\frac{1}{\pi}\operatorname{Im}%
\ln\det\left(  \underline{\underline{t}}_{c}^{\left(  \widehat{n}\right)
}\left(  \varepsilon\right)  ^{-1}-\underline{\underline{G}}_{0}\left(
\varepsilon\right)  \right)  \;, 
\end{equation}
is independent of the configuration, and can be written as an integral
over reciprocal wave-vector ${\bf k}$-space
\begin{equation}
N_{c}^{\left(  \widehat{n}\right)  }\left(  \varepsilon\right)  =
\frac{1}{\pi} \operatorname{Im} \int \ln\det \left( \underline{t}_c^{\left( 
\widehat{n}\right)}\left(  \varepsilon\right)  ^{-1}-
\underline{G}_0 \left( {\bf k},\varepsilon  \right) \right) d{\bf k}
\;, \label{Ncpa}
\end{equation}
while
\begin{equation}
\Delta N^{\left(  \widehat{n}\right)  }\left(  \varepsilon;\left\{
\widehat{e}\right\}  \right)  =-\frac{1}{\pi}\operatorname{Im}\ln\det\left(
\underline{\underline{I}}-\left(  \underline{\underline{t}}_{c}^{\left(
\widehat{n}\right)  }\left(  \varepsilon\right)  ^{-1}-\underline
{\underline{t}} \left(  \varepsilon;\left\{
\widehat{e}\right\}  \right)  ^{-1}\right)  \underline{\underline{\tau}}%
_{c}^{\left(  \widehat{n}\right)  }\left(  \varepsilon\right)  \right)  \;,
\end{equation}
is the only configuration dependent part of $N^{\left(  \widehat{n}\right)
}\left(  \varepsilon;\left\{  \widehat{e}\right\}  \right)  .$ Decomposing
$\underline{\underline{\tau}}_{c}^{\left(  \widehat{n}\right)  }\left(
\varepsilon\right)  $ into a site-diagonal, $\underline{\underline{\tau}}%
_{c}^{d\left(  \widehat{n}\right)  }\left(  \varepsilon\right)  $, and a
purely site-off-diagonal term, $\underline{\underline{\tau}}_{c}^{o\left(
\widehat{n}\right)  }\left(  \varepsilon\right)  $,%
\begin{equation}
\underline{\underline{\tau}}_{c}^{d\left(  \widehat{n}\right)  }\left(
\varepsilon\right)  =\left\{  \underline{\tau}_{c,ii}^{\left(  \widehat
{n}\right)  }\left(  \varepsilon\right)  \,\delta_{ij}\right\}
\;,\;\underline{\underline{\tau}}_{c}^{o\left(  \widehat{n}\right)  }\left(
\varepsilon\right)  =\left\{  \underline{\tau}_{c,ij}^{\left(  \widehat
{n}\right)  }\left(  \varepsilon\right)  \,\left(  1-\delta_{ij}\right)
\right\}  \;,
\end{equation}
$N^{\left(  \widehat{n}\right)  }\left(  \varepsilon;\left\{  \widehat
{e}\right\}  \right)  $ can further be evaluated as%
\begin{equation}
\Delta N^{\left(  \widehat{n}\right)  }\left(  \varepsilon;\left\{
\widehat{e}\right\}  \right)  =-\frac{1}{\pi}\operatorname{Im}\ln
\det\underline{\underline{\sl{M}}}^{\left(  \widehat{n}\right)  }\left(
\varepsilon;\left\{  \widehat{e}\right\}  \right)  -\frac{1}{\pi
}\operatorname{Im}\ln\det\left(  \underline{\underline{I}}-\underline
{\underline{X}}^{\left(  \widehat{n}\right)  }\left(  \varepsilon;\left\{
\widehat{e}\right\}  \right)  \underline{\underline{\tau}}_{c}^{o\left(
\widehat{n}\right)  }\left(  \varepsilon\right)  \right)  \;,
\end{equation}
where%
\begin{align}
\underline{\underline{\sl{M}}}^{\left(  \widehat{n}\right)  }\left(  \varepsilon
;\left\{  \widehat{e}\right\}  \right)   &  =\underline{\underline{I}}+\left(
\underline{\underline{t}} \left(
\varepsilon;\left\{  \widehat{e}\right\}\right)  ^{-1}-\underline{\underline{t}}_{c}^{\left(  \widehat
{n}\right)  }\left(  \varepsilon \right)
^{-1}\right)  \underline{\underline{\tau}}_{c}^{d\left(  \widehat{n}\right)
}\left(  \varepsilon\right) \\
&  =\left\{  \underline{\sl{M}}_{i}^{\left(  \widehat{n}\right)  }\left(
\varepsilon;\widehat{e}_{i}\right)  \,\delta_{ij}\right\}  \;,
\end{align}
with the matrices, $\underline{\sl{M}}_{i}^{\left(  \widehat{n}\right)  }\left(
\varepsilon;\widehat{e}_{i}\right)=\underline{D}_{i}^{\left(  \widehat{n}\right)  }\left(
\varepsilon,\widehat{e}_{i}\right)^{-1} $,defined in Eq. (\ref{Dmatrix}).
$
\underline{\underline{X}}^{\left(  \widehat{n}\right)  }\left(  \varepsilon
;\left\{  \widehat{e}\right\}  \right)   =
\left\{  \underline{X}_{i}^{\left(  \widehat{n}\right)  }\left(
\varepsilon;\widehat{e}_{i}\right)  \,\delta_{ij}\right\}$,
where $\underline{X}_{i}^{\left(  \widehat{n}\right)  }\left(
\varepsilon,\widehat{e}_{i}\right)  $, is defined in Eq. (\ref{Xmatrix}).
Therefore,
\begin{equation}
\Delta N^{\left(  \widehat{n}\right)  }\left(  \varepsilon;\left\{
\widehat{e}\right\}  \right)  =-\frac{1}{\pi}\sum_{i}\operatorname{Im}\ln
\det\underline{\sl{M}}_{i}^{\left(  \widehat{n}\right)  }\left(  \varepsilon
;\widehat{e}_{i}\right)  -\frac{1}{\pi}\operatorname{Im}\ln\det\left(
\underline{\underline{I}}-\underline{\underline{X}}^{\left(  \widehat
{n}\right)  }\left(  \varepsilon;\left\{  \widehat{e}\right\}  \right)
\underline{\underline{\tau}}_{c}^{o\left(  \widehat{n}\right)  }\left(
\varepsilon\right)  \right)  \;.
\end{equation}
Since, $\underline{\sl{M}}_{i}^{\left(  \widehat{n}\right)  }\left(  \varepsilon
;\widehat{e}_{i}\right)  $ depends by definition only on the orientation
$\widehat{e}_{i}$, the restricted average, $\left\langle \quad\right\rangle
_{\widehat{e}_{i}}$, of the first term simply equals to $
-\frac{1}{\pi}\operatorname{Im}\ln\det\underline{\sl{M}}_{i}^{\left(  \widehat
{n}\right)  }\left(  \varepsilon;\widehat{e}_{i}\right)$,
while the second term requires more care. Namely,
\begin{align}
&  \mathrm{Tr}\ln\left(  \underline{\underline{I}}-\underline{\underline{X}%
}^{\left(  \widehat{n}\right)  }\left(  \varepsilon;\left\{  \widehat
{e}\right\}  \right)  \underline{\underline{\tau}}_{c}^{o\left(  \widehat
{n}\right)  }\left(  \varepsilon\right)  \right)  =\sum_{k=1}^{\infty}\frac
{1}{k}\mathrm{Tr}\left(  \underline{\underline{X}}^{\left(  \widehat
{n}\right)  }\left(  \varepsilon;\left\{  \widehat{e}\right\}  \right)
\underline{\underline{\tau}}_{c}^{o\left(  \widehat{n}\right)  }\left(
\varepsilon\right)  \right)  ^{k}\\
&  =\sum_{k=2,4,\cdots}^{\infty}\frac{1}{k}\sum_{i_{1}\neq i_{2}\neq\ldots\neq
i_{k-1}\neq i_{k}}\mathrm{tr}\left(  \underline{X}_{i_{1}}^{\left(
\widehat{n}\right)  }\left(  \varepsilon;\widehat{e}_{i_{1}}\right)
\underline{\underline{\tau}}_{c,i_{i}i_{2}}^{\left(  \widehat{n}\right)
}\left(  \varepsilon\right)  \underline{X}_{i_{2}}^{\left(  \widehat
{n}\right)  }\left(  \varepsilon;\widehat{e}_{i_{2}}\right)  \ldots
\underline{X}_{i_{k}}^{\left(  \widehat{n}\right)  }\left(  \varepsilon
;\widehat{e}_{i_{k}}\right)  \underline{\underline{\tau}}_{c,i_{k}i_{1}%
}^{\left(  \widehat{n}\right)  }\left(  \varepsilon\right)  \right)  \;,
\end{align}
with tr$\left(  \quad\right)  $ denoting a trace in angular momentum 
space only.  From
the single-site CPA condition, Eq. (\ref{CPA-3}), it follows that the restricted
average of the term $k=2$ identically vanishes and that for higher
order $k$ terms the only elements in the sums which
contribute are those for which each of the
indices $i_{l}$ $\left(  l=1,\ldots,k\right)  $ occurs at least twice. 
These backscattering terms are neglected in the single-site CPA averages.
A useful property of the configurationally averaged integrated 
density of states given by the CPA
\begin{align}
\left\langle N^{\left(  \widehat
{n}\right)  }\left(  \varepsilon;\left\{  \widehat{e}\right\}  
\right) \right\rangle & = N_0\left(  \varepsilon\right)
-\frac{1}{\pi}\operatorname{Im}
\ln\det\left(  \underline{\underline{t}}_{c}^{\left(  \widehat{n}\right)
}\left(  \varepsilon\right)  ^{-1}-\underline{\underline{G}}_{0}\left(
\varepsilon\right)  \right) \nonumber \\
 &  -\frac{1}{\pi}\operatorname{Im} \sum_i \left\langle
\ln\det \underline{\sl{M}}_{i}^{\left(  \widehat
{n}\right)  }\left(  \varepsilon;\widehat{e}_{i}\right)\right\rangle
\label{Lloyd-cpa}
\end{align}
is that it is stationary with respect to changes in the t-matrices, 
$ \underline{\underline{t}}_{c}^{\left(  \widehat{n}\right)
}\left(  \varepsilon\right)$, which
determine the effective CPA medium. Indeed this stationarity condition 
can be shown to be another way of expressing the CPA 
condition~\cite{Faulkner+Stocks}. We will use this shortly in our derivation
of a robust expression for the calculation of the MCA.

The partially averaged electronic Grand Potential is given by
\begin{align}
\left\langle \Omega^{\left(  \widehat{n}\right)  }\right\rangle _{\widehat
{e}_{i}}= & -\int d\varepsilon\,f_{FD}\left(  \varepsilon
;\nu^{\left(  \widehat{n}\right)  }\right) N_{c}^{\left(  \widehat{n}\right)  }\left(  \varepsilon\right) + \frac{1}{\pi}\int d\varepsilon\,f_{FD}\left(  \varepsilon
;\nu^{\left(  \widehat{n}\right)  }\right)  \,\operatorname{Im}\ln
\det\underline{\sl{M}}_{i}^{\left(  \widehat{n}\right)  }\left(  \varepsilon
;\widehat{e}_{i}\right)  \;, \nonumber \\
 & + \sum_{j \ne i} \frac{1}{\pi}\int d\varepsilon\, f_{FD}
\left(  \varepsilon ;\nu^{\left(  \widehat{n}\right)  }\right)  \,
\operatorname{Im} \left \langle \ln \det\underline{M}_{j}^{\left(  
\widehat{n}\right)  }\left(  \varepsilon ;\widehat{e}_{j}\right) 
\right\rangle \;, \label{Omega}
\end{align}
and the Weiss field, $h_{i}^{\left(  \widehat{n}\right)  }$, can be 
expressed, using Eq.(\ref{WF-1}), as
\begin{equation}
h_{i}^{\left(  \widehat{n}\right)  }=\frac{3}{4\pi}\int\left(  \widehat{e}%
_{i}\cdot\widehat{n}\right)  \left[  \int d\varepsilon\,f_{FD}\left(
\varepsilon;\nu^{\left(  \widehat{n}\right)  }\right)  \frac{1}{\pi
}\operatorname{Im}\ln\det\underline{\sl{M}}_{i}^{\left(  \widehat{n}\right)
}\left(  \varepsilon;\widehat{e}_{i}\right)  \right]  \,d\widehat{e}_{i}\;.
\label{WF-2}%
\end{equation}

The solution of Eqs.(\ref{WF-2}) and (\ref{mag}) produces the variation 
of the magnetisation $m_{i}^{(\widehat{n})}$ with temperature $T$ with 
$m_{i}^{(\widehat{n})}$ going to zero at $T=T^{(\widehat{n})}_c$. 
When relativistic effects are included, the magnetisation direction
$\widehat{n}$ for which $T^{(\widehat{n})}_c$ is highest indicates the
easy direction for the onset of magnetic order. We can define a 
temperature range $\Delta T_{aniso}=
T^{(\widehat{n}_{e})}_c -T^{(\widehat{n}_{h})}_c$ where $\widehat{n}_{e}$
 and $\widehat{n}_{h}$ are the system's high temperature easy and hard
directions respectively, which is related to the 
magnetic anisotropy of the system at lower temperatures. Indeed
an adaptation of this approach to systems such as thin films 
 in combination with $T=0$K calculations may be useful in understanding 
temperature-induced spin reorientation transitions.~\cite{Laszlo}

\section{THEORETICAL FORMALISM FOR THE MAGNETISATION DEPENDENCE OF 
MAGNETIC ANISOTROPY AB-INITIO}

In the ferromagnetic state, at temperatures more than $\Delta T_{aniso}$
 below the Curie temperature, the magnetic anisotropy is given by the 
difference between the free energies, $F^{(\widehat{n})}$,
for two different magnetisation directions, $\widehat{n}_1$,
$\widehat{n}_2$, but the same magnetisation $m$ and therefore the
same values of the products of the Weiss field magnitudes with $\beta$,
i.e. $\beta h_{i}^{(\hat{n}_1)}=\beta h_{i}^{(\hat{n}_2)}$. 
Within our DLM theory
this means that the single site entropy terms in Eq.(\ref{FreeEnergy})
for each magnetisation direction cancel when the difference is taken
and the magnetic anisotropy energy MCA can be written
\begin{equation}
\Delta F(\widehat{n}_1,\widehat{n}_2) =  \left\langle \Omega^{\left(  
\widehat{n}_1\right)  }\right\rangle - \left\langle \Omega^{\left(  
\widehat{n}_2\right)  }\right\rangle.
\end{equation}
This becomes $
\Delta F(\widehat{n}_1,\widehat{n}_2) \approx -\int d\varepsilon\,
f_{FD}\left(  \varepsilon ;\nu^{\left(  \widehat{n}_1\right)  }\right)
\left( \left\langle N^{\left(  \widehat{n}_1 \right)  }\left(  
\varepsilon\right)\right\rangle - \left\langle
N^{\left(  \widehat{n}_2 \right)  }\left(  \varepsilon\right) 
\right\rangle \right)$
where a small approximation is made through the use of the one 
chemical potential $\nu^{\left(  \widehat{n}_1\right)  }$.
Using Eqs.(\ref{Lloyd-cpa}) and (\ref{Ncpa}) this can be written 
explicitly as 
\begin{align}
\Delta F(\widehat{n}_1,\widehat{n}_2) &  = -\int d\varepsilon\, 
f_{FD}\left(  \varepsilon ;\nu^{\left(  \widehat{n}_1\right)  }\right)
\frac{\operatorname{Im}}{\pi} \int
\ln\det\left(\underline{1}+ \left[  
\underline{\underline{t}}_{c}^{\left(  \widehat{n}_1 \right)
}\left(  \varepsilon\right)^{-1}- 
\underline{\underline{t}}_{c}^{\left(  \widehat{n}_2 \right)
}\left(  \varepsilon\right)^{-1} \right] \underline{\tau}_c^{\hat{n}_2}(\varepsilon,{\bf k})
  \right) d{\bf k} \nonumber \\
-\sum_{i} \int d\varepsilon\, f_{FD} & \left(  \varepsilon ; 
\nu^{\left(\widehat{n}_1\right)  }\right) \frac{\operatorname{Im}}{\pi} 
\int  \left[ P_{i}^{\left(  \widehat{n}_1 \right)  }\left(  
\widehat{e}_{i}\right)
\ln\det\underline{M}_{i}^{\left(  \widehat{n}_1 \right)
}\left(  \varepsilon;\widehat{e}_{i} \right)- P_{i}^{\left(  
\widehat{n}_2\right)  }\left(  \widehat{e}_{i} \right) 
\ln\det\underline{M}_{i}^{\left(  \widehat{n}_2 \right)
}\left(  \varepsilon;\widehat{e}_{i} \right) \right]
d\widehat{e}_{i}
\end{align}

As well as ensuring that the Brillouin zone integration in the above
equation is accomplished with high numerical 
precision~\cite{Razee+97,Razee+99},
care must also be taken to establish accurately the two CPA media describing 
the system magnetised along the two directions, $\widehat{n}_1$ and 
$\widehat{n}_2$. Eq.(\ref{CPA-3}) has to be solved to high precision in
each case. These steps were successfully taken and tested for our first 
application on $L1_0$ $FePt$~\cite{MAEvsT}. We have found however that a less computationally demanding scheme for extracting the temperature
dependence of the magnetic anisotropy can be derived by consideration
of the magnetic torque. It is sufficiently robust numerically to be
applicable to a range of magnetic materials, whether hard or 
soft magnetically and in bulk, film and nanoparticulate form.  

\subsection{A TORQUE-BASED FORMULA FOR THE MAGNETISATION DEPENDENCE OF
MAGNETIC ANISOTROPY}

We return again to Eq.(\ref{FreeEnergy}) for the expression for the
free energy $F^{\left( \widehat{n} \right)}$ of a system magnetised 
along a direction $\widehat{n} =$ ( $\sin \vartheta \cos \varphi$,
$\sin \vartheta \sin \varphi $, $\cos \vartheta$ ) and consider how it 
varies with change in magnetisation angles $\vartheta$ and $\varphi$,
i.e. $T_{\vartheta} = -\frac{\partial F^{\left( \widehat{n} 
\right)}}{\partial \vartheta}$, $T_{\varphi} = 
-\frac{\partial F^{\left( \widehat{n} \right)}}{\partial \varphi}$. Since
the single site entropy term in Eq.(\ref{FreeEnergy}) is invariant with
respect to the angular variations we can write
\begin{equation}
T_{\vartheta (\varphi)}= -\frac{\partial}{\partial \vartheta (\varphi)} 
\left[ \sum_i \int P_{i}^{\left(  \widehat{n}\right)  }\left(  
\widehat{e}_i \right) \left\langle \Omega^{\left(  
\widehat{n}\right)  }\right\rangle _{\widehat{e}_i} d\widehat{e}_i
\right].
\end{equation}
By using Eqs.(\ref{Lloyd-cpa}) and (\ref{Omega}),
together with the 
stationarity of the CPA integrated density of states to variations of
the CPA effective medium, we can write directly
\begin{equation}
T_{\vartheta (\varphi)}=-\frac{\operatorname{Im}}{\pi}\int d\varepsilon\,f_{FD}\left(  \varepsilon
;\nu^{\left(  \widehat{n}\right)  }\right)  \,\left[ \sum_i \int 
\frac{\partial P_{i}^{\left(  \widehat{n}\right)}\left(
\widehat{e}_i \right)  }{\partial \vartheta (\varphi)}
\ln \det\underline{M}_{i}^{\left(  \widehat{n}\right)  }\left(  
\varepsilon;\widehat{e}_i \right) d\widehat{e}_i \right] \label{torque}
\end{equation}
According to the form of $ P_{i}^{\left(  \widehat{n}\right)}\left(
\widehat{e}_i \right)$ given in Eq. (\ref{Pi}) the principal expression for the magnetic torque at finite temperature
is thus
\begin{equation}
T_{\vartheta(\varphi)}=\frac{\operatorname{Im}}{\pi} \int d\varepsilon\,
f_{FD}\left(  \varepsilon ;\nu^{\left(  \widehat{n}\right)  }\right)  \,
\left[ \sum_{i}
\int \beta h_i P_{i}^{\left( \widehat{n}\right) }\left(\widehat{e}_i \right)
 \left( \frac{\partial \widehat{n}}{
\partial \vartheta(\varphi)} \cdot \widehat{e}_i \right) \,
\ln \det\underline{M}_{i}^{\left(  \widehat{n}\right)  }\left(
\varepsilon;\widehat{e}_i \right) d\widehat{e}_i \right] \label{torque2}.
\end{equation} 

For a uniaxial ferromagnet such as a $L1_0$ 3d-4d/5d transition metal 
magnet or a magnetic thin film, the performance of a single CPA calculation for
appropriate values of the energy $ \varepsilon$ is carried out at fixed values of 
the $\beta h_i$ products (and therefore a chosen magnetisation $m$) and for the system magnetised along $\widehat{n}= (\sin \pi/4,0,\cos \pi/4)$.  Subsequent evaluation of our torque expression, Eq.(\ref{torque2}), 
i.e  $-T_{\vartheta} ( \vartheta=\pi/4, \varphi=0)$ yields the sum of 
the first two magnetic anisotropy constants $K_2$ and $K_4$. Similarly
$-T_{\varphi} ( \vartheta=\pi/2, \varphi=\pi/8)$ gives an estimate of
the leading constant $K_{1}/2$ for a cubic system. In the appendix we
derive $T_{\vartheta (\varphi)}$ for a magnet at $T=0$K and show that
this is equivalent to Eq.(\ref{torque}) for the limit 
$\beta h \rightarrow \infty$, i.e. when $T \rightarrow 0$K.

\section{THE CALCULATION OF THE TEMPERATURE DEPENDENCE OF THE MAGNETISATION, $M(T)$, AND THE $M(T)$ DEPENDENCE OF THE MAGNETIC ANISOTROPY}

In a first-principles implementation of the DLM picture,
the averaging over the orientational configurations of the
local moments is performed using techniques
adopted from the theory of random metallic
alloys.~\cite{scf-kkr-cpa,DLM} Over the past 20 years,
the paramagnetic state, onset of magnetic order and transition
temperatures of many systems have been successfully
described. All applications to date, apart from our
earlier study of $FePt$~\cite{MAEvsT} and the cases described in this paper, 
have, however, neglected relativistic effects and have been devoted
to the paramagnetic state where the symmetry turns the calculation into a 
binary alloy-type one with
half the moments oriented along a direction and the rest antiparallel.
Once relativistic effects are included and/or the ferromagnetic state is
considered, this simplicity is lost and, as is shown above, the continuous probability 
distribution, $P_{i}^{(\widehat{n})} (\widehat{e}_i)$, must be sampled 
for a fine mesh of angles
and the averages with the probability distribution performed numerically.
(Careful checks have to be made to ensure that the sampling of
$P_{i}^{(\widehat{n})} (\widehat{e}_i)$ is sufficient - in our calculations
up to 40,000 values are used.)
Of course, in the paramagnetic state $P_{i}^{(\widehat{n})} (\widehat{e}_i) = 
\frac{1}{4 \pi}$
so that a local moment on a site has an equal probability in pointing
in any direction $\widehat{e}_i$.

 The local moments 
change their orientations, $\{ \widehat{e}_{i} \}$, on a time scale
$\tau$ 
long in comparison with the time taken for electrons to `hop' from
site to site.  Meanwhile
their magnitudes fluctuate rapidly on this fast electronic 
time scale which means that over times $\tau$,
the magnetisation on a site is equal to $\mu_{i} 
\widehat{e}_{i}$. 
As a consequence of the itinerant nature of the electrons, the 
magnitude $\mu_{i}$ depends on the orientations of the 
local moments on surrounding sites, i.e. $\mu_{i}= \mu_{i}(\{ 
\hat{e}_{l} \})$. In the DLM theory described above,
$\mu_{i} = \langle \mu_{i}(\{\hat{e} \}) 
\rangle_{\widehat{e}_{i}}$, so that the size of the local moment on a 
site is taken from  
 electronic charge density spin-polarised along $\widehat{e}_{i}$ and
integrated over the site. An average is taken over the orientations
$\{\hat{e} \}$ on surrounding sites and the local charge and
magnetisation densities are calculated self-consistently from a
generalised SDFT formalism and SCF-KKR-CPA techniques. 

Being a local mean field 
theory, the principal failure of the DLM to date
is that it does not give an adequate description of local moment
formation in Ni rich systems because it cannot allow for the effects 
of correlations among the orientations of the local moments
over small neighbourhoods of atomic sites. (In principle this 
shortcoming is now addressable using the newly developed SCF-KKR-NLCPA 
method~\cite{Rowlands,NLCPA-totE}.) In this paper however we will focus entirely on 
`good' local
moment systems where the sizes of the moments are 
rather insensitive to their orientational environments. In these cases,
 for example, the self-consistently determined local moments of the 
paramagnetic DLM state differ little from the magnetisation per site 
obtained for the ferromagnetic state. For example in paramagnetic DLM
$L1_0$-FePd, a local moment of 2.98 $\mu_B$ is set up on each $Fe$ site whilst 
no moment forms on
the $Pd$ sites. For the same lattice spacings ($c=$0.381nm, $c/a =$1 - note we have neglected the
deviation of $c/a$ from ideal found experimentally) we find that,
for the completely ferromagnetically ordered state of $FePd$ at $T=0$K, the 
magnetisation per $Fe$ site is 2.96$\mu_B$ and a small magnetisation of 0.32$\mu_B$
is associated with the $Pd$ sites. Likewise, the $Fe_{50}Pt_{50}$ f.c.c. 
disordered alloy has local moments of 2.92$\mu_B$ on the $Fe$ sites in the paramagnetic
state whilst the ferromagnetic state has magnetisation of 2.93$\mu_B$ and 0.22$\mu_B$
 on each $Fe$ and $Pt$ site respectively ($a=$0.385nm). We can therefore safely use the 
self-consistently generated
effective potentials and magnetic fields for the paramagnetic DLM state along with the charge and magnetisation densities for calculations for the ferromagnetic state below $T_c$.

Our calculational method therefore is comprised of the following steps.
\begin{itemize}
\item[1.] Perform self-consistent scalar-relativistic DLM calculations
 for the paramagnetic state, $T > T_c$, to form effective potentials 
and magnetic fields from the local charge and magnetisation densities
(using typically the local spin density approximation, (LSDA)).
 This fixes
the single-site \emph{t}-matrices, $\underline{t}_{i}^{\left(  \widehat
{n}\right)  }\left(  \widehat{z}\right)  $

\item[2.] For a given temperature and orientation, 
$\widehat{n}= (\sin \vartheta \cos \varphi,\sin \vartheta \sin \varphi,
\cos \vartheta)$
determine the $h_{i}^{\left(  \widehat{n}\right)  }$'s ( and also the
chemical potential $\nu^{\left(  \widehat{n}\right)  }$ from 
Eq.(\ref{Lloyd-cpa})) selfconsistently:

\begin{enumerate}
\item[(a)] for a set of $\lambda_{i}=\beta h_{i}^{\left(  \widehat{n}\right)
}$ determine the $\underline{t}_{i,c}^{\left(  \widehat{n}\right)  }$ 
by solving the CPA condition, Eq. (\ref{CPA-3});

\item[(b)] calculate new Weiss fields, Eq. (\ref{WF-2});

\item[(c)] repeat steps 2.(a) and (b) until 
convergence. [For a system where there is a single local moment per 
unit cell, this iterative procedure can be circumvented. A 
series of values of $\lambda(=\beta h^{(\widehat{n})})$ is picked to 
set the probabilities,
$P^{(\widehat{n})} (\widehat{e}_i)$ (and magnetisations $\vec{m}= \langle \widehat{e} \rangle $, $m= |\vec{m}|$). The Weiss field
$ h^{\left(  \widehat{n}\right)}$is then calculated from (\ref{WF-2})
and the ratio of $h^{\hat{n}}$ to $\lambda$ then uniquely determines 
the temperature $T$ for each of the initially chosen values of 
$\lambda$ and hence the temperature dependence of the magnetisation.]
\end{enumerate}

\item[3.] Calculate the torque, $T_{ \vartheta(\varphi)}$  from 
Eq.(\ref{torque2}) to give the magnetic anisotropy and also average 
alignment of the local moments,
 $\vec{m}_{i}^{\left(  \widehat{n}\right)  }\left(  
T\right)  $, proportional to the total magnetisation, $ \mu_i \vec{m}_{i}$, from Eq. (\ref{mag}).

\item[4.] Repeat steps 2. and 3. for a different direction,
$\widehat{n}^{\prime}$ if necessary. 
\end{itemize}
(On a technical point: all integrals over energy $\frac{\operatorname{Im}}{\pi} \int d\varepsilon\,
f_{FD}\left(  \varepsilon ;\nu^{\left(  \widehat{n}\right)  }\right) 
\cdots$ are carried out via a suitable contour in the complex energy plane and a summation over Matsubara frequencies~\cite{dynsusc-bigpaper}. We
 use a simple box contour which encloses Matsubara frequencies up to
$\approx$ 10 eV$/\hbar$.) 

In the following examples for the uniaxial
ferromagnets $FePt$ and $FePd$, we have carried through 
steps 1. to 4. for $\vartheta_1 = \frac{\pi}{4}$,$\varphi_1 = 0$ 
where $T_{\vartheta} = -(K_2 +K_4)$, ($T_{\varphi}=0$) and also for 
$\vartheta_2=\frac{\pi}{3}$, $\varphi_2 = 0$ where $T_{\vartheta} =
 -\frac{\sqrt{3}}{2}(K_2+\frac{3}{2} K_4)$, ($T_{\varphi}=0$). For 
the cubic magnet disordered
$FePt$ we use $\vartheta_1 = \frac{\pi}{4}$,$\varphi_1 = 0$ as a 
numerical check, where both $T_{ \vartheta}$ and $T_{ \varphi}$ should
 be zero, and $\vartheta_2= \frac{\pi}{2}$,$\varphi_2=\frac{\pi}{8}$
where $T_{ \vartheta}=0$ and $T_{ \varphi}= - \frac{K_1}{2}$.

\section{UNIAXIAL MAGNETIC ANISOTROPY - FERROMAGNETS WITH TETRAGONAL CRYSTAL SYMMETRY}

Our case study in this paper is $L1_0$-FePd and the trends we find are very similar
to those found for $L1_0$-FePt. We carry out the steps 1-4 elaborated
in the last section. Figure 1 shows the dependence of the magnetisation upon 
temperature. In this mean field approximation we find a Curie temperature of
1105K in reasonable agreement with the experimental value of 723K~\cite{TcFePd}. (An Onsager 
cavity field technique could be used to improve this
estimate, see ~\cite{JBS+BLG}, without affecting the quality of the following 
results for $K$.) Although the
shortcomings of the mean field approach do not produce the spinwave $T^{
\frac{3}{2}}$ behavior at low temperatures, the easy axis for the onset of
magnetic order is deduced, $\widehat{n}=(0,0,1)$ perpendicular to the layering of the $Fe$ and 
$Pd$ atoms, (not shown
in the figure) and it corresponds to that found at lower temperatures
both experimentally~\cite{FePd-expt} and in all theoretical ($T=0$K) calculations~\cite{FePd-theory}. 
\begin{figure}[tbh]
\resizebox{0.9\columnwidth}{!}{\includegraphics*{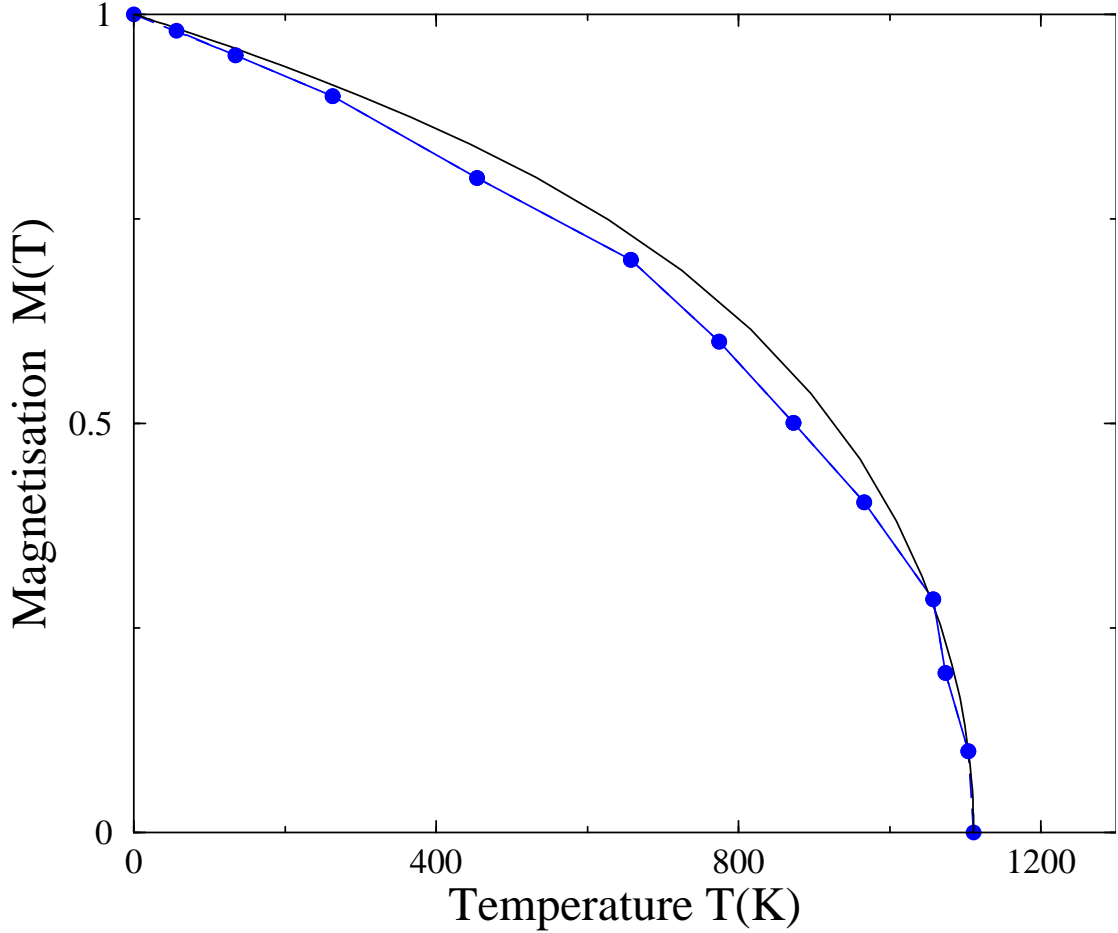}}
\caption{The magnetisation of FePt versus temperature. The
filled circles refer to a magnetisation along $\hat{n}= (1,0,1)$.
$T_c$ is at 1105K with the easy axis, $(0,0,1)$. The full line shows the mean field approximation to a classical Heisenberg
model for comparison.}
\end{figure}
Figure 2 shows the magnetic anisotropy energy, $\Delta F ((0,0,1),(1,0,0))=
-(K_2+K_4)$ versus the square of the magnetisation. The same linear relationship that we
found for $FePt$~\cite{MAEvsT} is evident, a clear consequence of the itinerant nature of the
 magnetism is this system. This magnetisation dependence 
differs significantly from that produced by the single ion model, also shown in
the figure. At $T=0$K, $K_2+K_4$ is 0.335meV is in fair agreement with the value of 0.45 meV inferred
from low temperature measurements on well ordered samples~\cite{FePd-expt} (as with $FePt$, $K$ decreases
significantly if the degree of long-range chemical order is reduced). The value is also in line with values of 0.1  to 0.5 meV found by other ab-initio approaches ~\cite{FePd-theory}.
\begin{figure}[tbh]
\resizebox{0.9\columnwidth}{!}{\includegraphics*{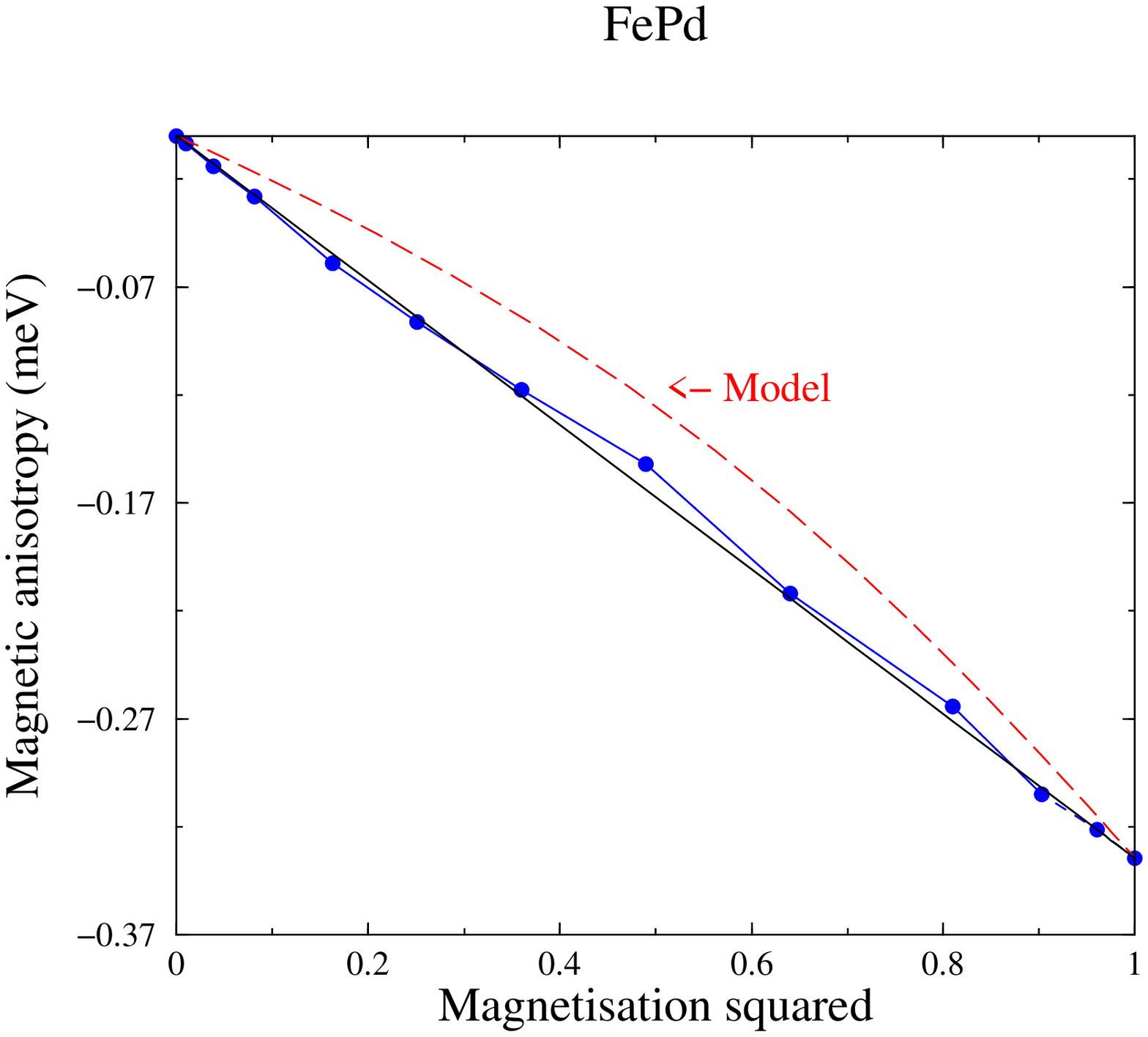}}
\caption{The magnetic anisotropy of FePd as a function of the square of
magnetisation. The filled circles show the calculations from the ab-initio theory,
 the full line $K_0 (m(T)/m(0))^2$ and the dashed line the single-ion model
function $K_0 < g_2 (\hat{e})>_T /  < g_2 (\hat{e})>_0$ with $K_0=$ -0.335meV.}
\end{figure}
\begin{figure}[tbh]
\resizebox{0.9\columnwidth}{!}{\includegraphics*{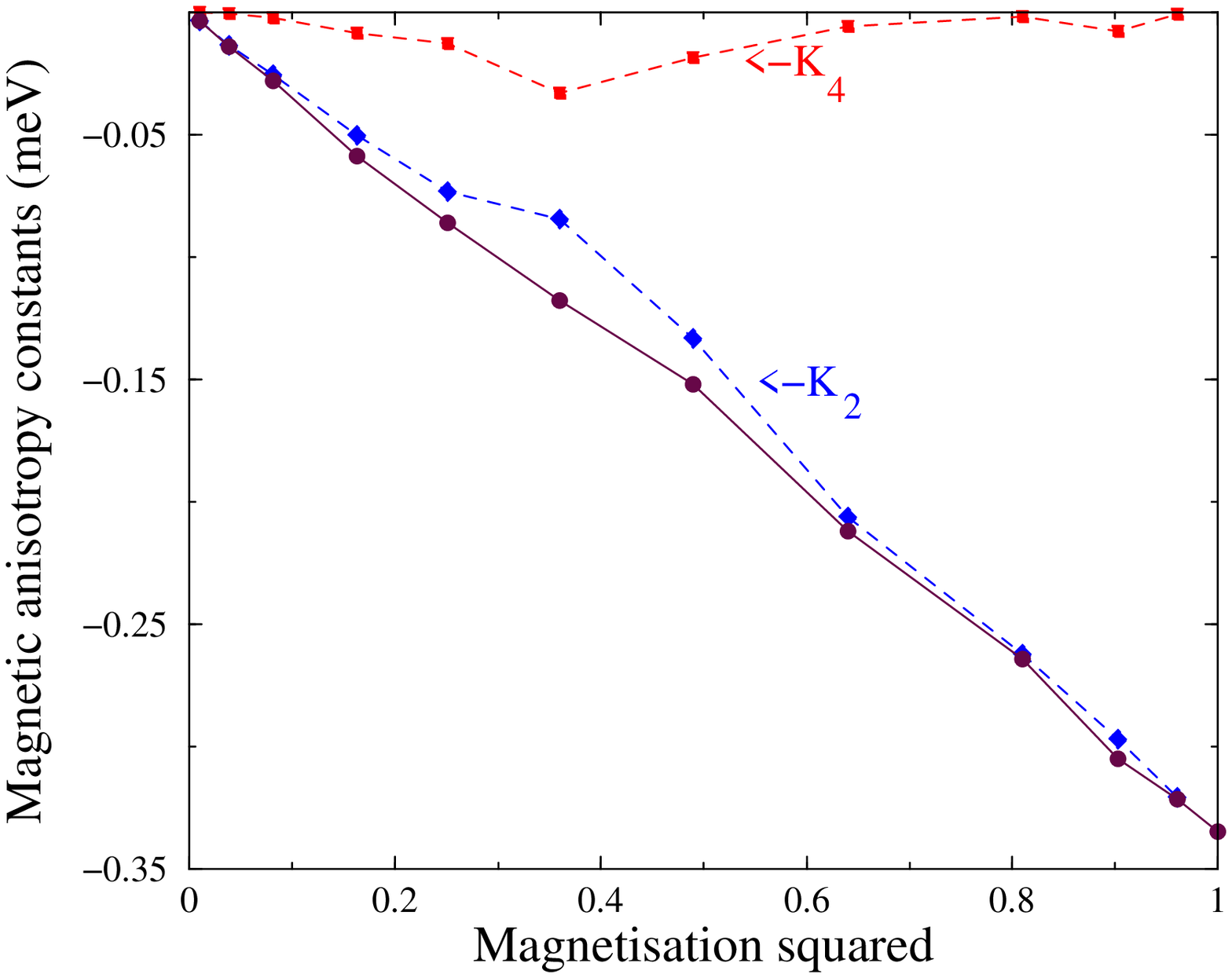}}
\caption{The magnetic anisotropy constants $K_2$, $K_4$ of FePd as a function of the square of
magnetisation. The filled circles and full line show the calculations from the ab-initio theory
of the sum, the dashed line with filled diamonds describes $K_2$ and the dotted line with squares shows $K_4$.}
\end{figure}
From $T_{\vartheta}$ for both $\vartheta=\frac{\pi}{4}$,$\varphi = 0$ 
and $\vartheta=\frac{\pi}{3}$,$\varphi = 0$ the magnitudes of the MCA constants 
$K_2$ and $K_4$ are extracted and shown in Figure 3. The dominance of $K_2$ is obvious but
it is also clear that the $m^2$ dependence is followed closely by the total anisotropy,
$K_2 + K_4$, and only approximately by the leading constant $K_2$. It is interesting to note
that an anisotropic classical Heisenberg model leads to similar $m$ dependence to
$K$ if treated within a mean field approach. To illustrate this point we show in Figure 4 the results of mean field calculations
of $K$ for a model with both single-ion and anisotropic nearest neighbor exchange,i.e. where the following hamiltonian
is appropriate:
\begin{equation}
H= -\frac{1}{2} \sum_{i,j} (J^{\|} (e_{x,i} e_{x,j}+ e_{y,i} e_{y,j})+ J^{\bot} e_{z,i} e_{z,j}) -k\sum_i (e_{z,i})^2
\end{equation}
The full curve shows the single ion model results for the limit  $J^{\|} = J^{\bot}$, which are
 also shown in Figure 3. At low $T$ as $m(T) \rightarrow 1$, $K(T)/K(0)$ has the familiar $m^{l (l+1)/2}$
 form with $l=1$ for a uniaxial magnet. By introducing a small difference between $J^{\|}$ and 
$J^{\bot}$, so that $J^{\bot}-J^{\|} =0.01 J^{\bot}$, $K(T)/K(0)$ varies as $m^2$.
\begin{figure}[tbh]
\resizebox{0.9\columnwidth}{!}{\includegraphics*{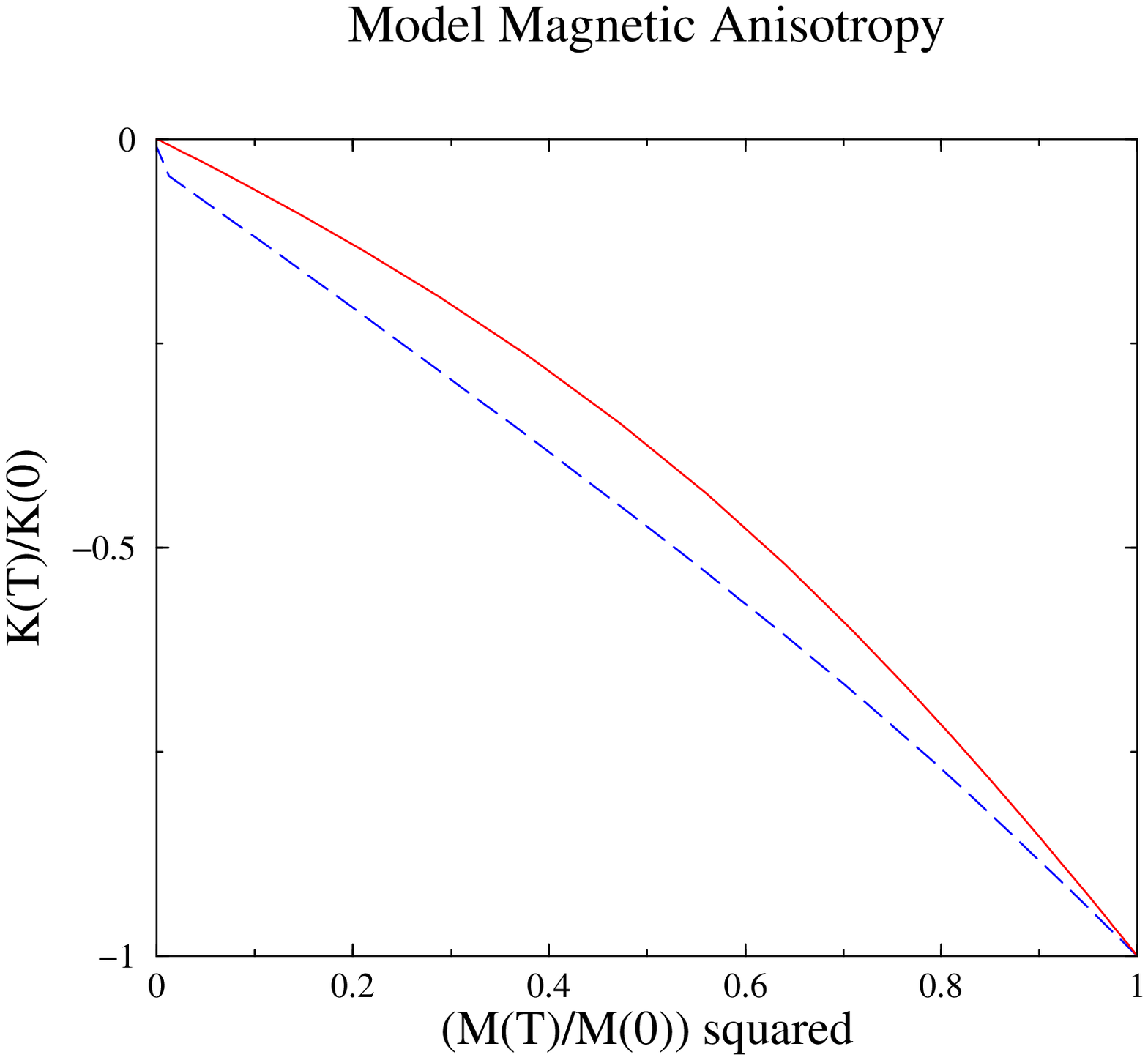}}
\caption{The magnetic anisotropy energy $K$ calculated in a mean field approximation for a model of a uniaxial magnet which has both anisotropic exchange and single ion anisotropy. The full line shows results
with single ion anisotropy only, $k= 0.002 J^{\bot}$ and $J^{\bot}-J^{\|} = 0$. The dashed line shows 
results for the same $k$ and $J^{\bot}-J^{\|} = 0.01 J^{\bot}$.}
\end{figure}

\section{CUBIC MAGNETIC ANISOTROPY - THE F.C.C. $Fe_{50}Pt_{50}$ SOLID SOLUTION}

Crystal structure is known to have a profound effect upon the magnetic anisotropy.
Magnetic anisotropy within a single ion anisotropy model decreases according to $m^{l(l+1)/2}$ 
at low $T$, ($m \approx 1$) and proportional to $m^l$ for small $m$ at higher $T$. For materials with tetragonal symmetry, $l=2$ as shown in Fig.2. On this basis a cubic magnet's $K$ should possess an $m$ dependence where $l=4$, i.e. $m^{10}$ at low $T$ and $m^4$ at 
higher $T$. In this section we show our results for the itinerant magnet, compositionally disordered $Fe_{50}Pt_{50}$.  In this system the lattice sites of the f.c.c. lattice are occupied at random by either $Fe$ or $Pt$ atoms. The cubic symmetry causes this alloy to be magnetically very soft. Ordering into a tetragonal $L1_0$ structure of layers of predominantly $Fe$ atoms stacked alternately with $Pt$ layers along the $(1,0,0)$ direction causes a significant increase of $K$. Okamoto et al~\cite{Okamoto} have measured $K$ of $FePt$ carefully as a function of compositional order and the trend, for $T=0$K, has been successfully reproduced in ab-initio calculations~\cite{ICNDR,Burkert}. 

As with our earlier calculations for $L1_0$-FePt~\cite{MAEvsT} and FePd, this disordered
alloy's magnetisation follows a similar T-dependence to that of a mean field treatment
of a classical Heisenberg model. We find a Curie temperature of 1085K, again a mean field value which 
is in reasonable agreement with the experimental value of 750K~\cite{Okamoto}.
Figure 5 shows our calculations of the magnetisation dependence of the leading 
magnetic anisotropy constant $K_1$ (Eq.\ref{MAE-cubic}). At $T=0$K, $\Delta F((0,0,1),
(1,1,1)) = -K_1/3$ is just 2.8$\mu$eV ($\pm 0.1 \mu$eV) , some three orders of magnitude smaller than the uniaxial MCA ($K_2+K_4$) we find its $L1_0$-ordered counterpart~\cite{MAEvsT}. Despite this small value we find that our method is robust enough to follow
the magnetisation and $T$-dependence of $K_1$. $K_1$ is determined from a calculation of 
$T_{\varphi}$ where for $\vartheta=\pi/2$ and $\varphi= \pi/8$ it equals $K_1/2$. 
As expected $K_1$ decreases rapidly - Fig. 5 depicts $K_1$ versus the fourth
power of the magnetisation. At low $T$ $K_1$ varies approximately as $m^{7}$ whereas this dependence
becomes $m^{4}$ for smaller $M$ and higher $T$. Fig.5 also shows the behavior of the
single ion model for a cubic system for comparison.  As with the uniaxial metallic magnets already investigated, the ab-initio R-DLM results differ significantly. 
\begin{figure}[tbh]
\resizebox{0.9\columnwidth}{!}{\includegraphics*{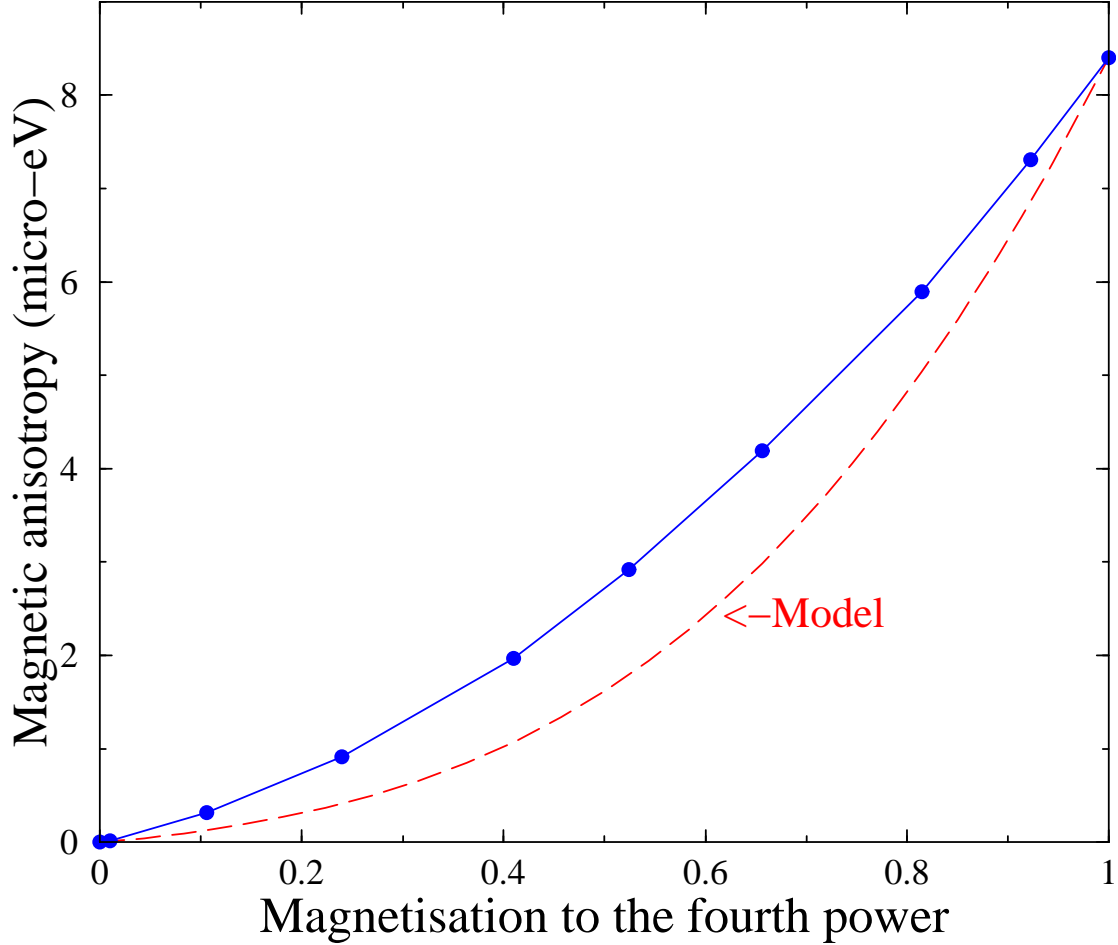}}
\caption{The magnetic anisotropy constant $K_1$ of the cubic magnet $Fe_{50}Pt_{50}$  as a function of the fourth power of the magnetisation, $m^4$. The filled circles show the
 calculations from the  ab-initio theory and the dashed
 line the single-ion model function $K_1^0 < g_2 (\hat{e})>_T /  < g_2 (\hat{e})>_0$ with
 $K_1^0=$ 8.4$\mu$eV.}
\end{figure}

\section{CONCLUSIONS}

We have shown that by including relativistic effects such as spin-orbit coupling
into the Disordered Local Moment theory of finite temperature magnetism, the temperature dependence of magnetic anisotropy can be obtained. Magnetic anisotropy is determined via consideration of magnetic torque expressed within a multiple-scattering formalism.  For uniaxial metallic magnets with tetragonal crystal symmetry, $L1_0$-$FePt$ and $FePd$,  we find $K$ to vary with the square of the overall magnetisation, $m(T)$. This is at odds with what an analysis based on a single ion anisotropy model would find but in agreement with experimental measurements for $FePt$. An interpretation in terms of an anisotropic Heisenberg model explains this behavior~\cite{Mryasov}. We suggest that this $m^2$ behavior is typical for high $T_c$ transition metal alloys ordered into a tetragonal structure. We find the first anisotropy coefficient, $K_2$ to be dominant. We have also investigated the magnetic anisotropy of metallic magnets with cubic 
crystal symmetry which are very soft magnetically. In the example of the f.c.c. 
substitutional alloy, $Fe_{50}Pt_{50}$, the leading constant $K_1$ decreases according to $m^n$ where $n$ ranges between 7 and 4 as the temperature is increased. This behavior also differs significantly from that of a simple single ion model. 
Application of this R-DLM theory of magnetism at finite temperature has been confined here
to bulk crystalline systems. It also, however, has particular relevance for thin film and nanostructured metallic magnets \cite{SKKR, Antoniak,Ebert}  where it can be used to uncover temperature-induced reorientation transitions. e.g. Buruzs et al. \cite{Laszlo} have recently applied the theory to $Fe$ and $Co$ monolayers on $Cu(111)$.  Future possible applications also include the study of the temperature dependence of magnetostriction, the design of high permeability materials and magnetotransport phenomena in spintronics.

\section{Acknowledgements}
We acknowledge support from the EPSRC(U.K), the Centre for Scientific Computing at the University of Warwick, the Hungarian National Science Foundation (OKTA T046267) and to the Center for Computational
Materials Science (Contract No. FWF W004 and GZ 45.547).

\appendix{APPENDIX A: TORQUE FOR $T \rightarrow 0$K}

Concerning the MCA of a ferromagnet at $T=0$K, the relevant part of the 
total energy is
\begin{equation}
F^{(\widehat{n})} = - \int d\varepsilon\,f_{FD}\left(
\varepsilon;\nu^{\left(
\widehat{n}\right)  }\right)  \,N^{\left(  \widehat{n}\right)  }\left(
\varepsilon \right)
\end{equation}
where $N^{\left(  \widehat{n}\right)  }\left(\varepsilon \right)$ is the 
integrated density of states, 
\begin{equation}
N^{\left(  \widehat{n}\right)  }\left(\varepsilon \right) = 
N_0 \left(\varepsilon \right) -\frac{1}{\pi
}\operatorname{Im}\ln\det\left(  \underline{\underline{t}} \left(
\widehat{n} ; \varepsilon
\right)  ^{-1}-\underline{\underline{G}}_{0}\left(  \varepsilon\right)
\right), 
\end{equation}
and the inverse of the single site t-matrix is
 \begin{equation}
\underline{t}_i \left( \widehat{n}; \varepsilon \right)=
\underline{R}\left(  \widehat{n} \right)  \,\underline{t}%
_{i} \left(  \widehat{z}; \varepsilon \right)  \underline
{R}\left(  \widehat{n} \right)  ^{+}.
\end{equation}
Now $\underline{R}\left(  \widehat{n} \right)= 
\exp  i \alpha_{\widehat{m}} (\widehat{m}\cdot \vec{\underline{J}})$ 
where $\alpha_{\widehat{m}}$ is the angle of rotation about an axis 
$\widehat{m} = (\widehat{z} \wedge \widehat{n})/|\widehat{z} \wedge 
\widehat{n}|$ and $\vec{\underline{J}}$ is the total angular momentum. 
The torque quantity $T_{\alpha_{\widehat{u}}}^{\left( 
\widehat{n}\right)  }= -\frac{\partial F^{(\widehat{n})}}{\partial \alpha_{\widehat{u}}}$, describing the variation of the
total energy with respect to a rotation of the magnetisation about an
axis $\widehat{u}$, is 
\begin{equation}
T_{\alpha_{\widehat{u}}}^{\left( \widehat{n}\right)  }= -\frac{1}{\pi} \int d\varepsilon\,f_{FD}\left(
\varepsilon;\nu^{\left( \widehat{n}\right)  }\right) \operatorname{Im}
\frac{\partial}{\partial \alpha_{\widehat{u}}} \left[ \ln\det\left(  \underline{\underline{t}} \left(
\widehat{n} ; \varepsilon
\right)  ^{-1}-\underline{\underline{G}}_{0}\left(  \varepsilon\right)
\right) \right]
\end{equation}
which can be written
\begin{equation}
T_{\alpha_{\widehat{u}}}^{\left( \widehat{n}\right)  }= 
-\frac{1}{\pi} \int d\varepsilon\, f_{FD}\left(\varepsilon;\nu^{\left( 
\widehat{n}\right)  }\right) \operatorname{Im} \sum_i tr
\left( \underline{\tau}^{\left( \widehat{n}\right)  }_{ii} \left( 
\varepsilon \right)
\frac{\partial}{\partial \alpha_{\widehat{u}}} \left( \underline{R}
\left( \widehat{n}\right) \underline{t}\left(\widehat{n} ; \varepsilon
\right)  ^{-1} \underline{R}\left(  \widehat{n} \right)^{+} \right)
\right).\label{T0}
\end{equation}
Since $\frac{\partial \underline{R}\left(  \widehat{n} 
\right)}{\partial \alpha_{\widehat{u}}} = i (\vec{\underline{J}} 
\cdot \widehat{u}) \underline{R}\left(  \widehat{n} \right)$ and 
$\frac{\partial \underline{R}\left(  \widehat{n} \right )^{+}}{\partial 
\alpha_{\widehat{u}}} = - i (\vec{\underline{J}} \cdot \widehat{u}) 
\underline{R}\left(  \widehat{n} \right)$,
\begin{equation}
T_{\alpha_{\widehat{u}}}^{\left( \widehat{n}\right)  }=\frac{1}{\pi} \int d\varepsilon\,f_{FD}\left(
\varepsilon;\nu^{\left( \widehat{n}\right)  }\right) \operatorname{Im} 
i \, \sum_i
tr \left( \underline{\tau}^{\left( \widehat{n}\right)  }_{ii}
\left( \varepsilon \right)  \left[ 
(\vec{\underline{J}} \cdot \widehat{u}) \underline{t}\left(\widehat{n} ; \varepsilon
\right)  ^{-1} - \underline{t}\left(\widehat{n} ; \varepsilon
\right)  ^{-1} (\vec{\underline{J}} \cdot \widehat{u}) \right] \right).
\end{equation}
For $T_{\vartheta (\varphi)}^{\left( \widehat{n}\right)  }$, 
$(\vec{\underline{J}} \cdot 
\widehat{u})$ is just $\underline{J}_{y(z)}$.

Consider now our finite temperature torque expression, Eq.(\ref{torque}), i.e. 
\begin{equation}
T_{\alpha}=\frac{\operatorname{Im}}{\pi}\int d\varepsilon\,f_{FD}\left(
\varepsilon;\nu^{\left(  \widehat{n}\right)  }\right)  \sum_{i}\int
\frac{\partial P_{i}^{\left(  \widehat{n}\right)  }\left(  \widehat{e}%
_{i}\right)  }{\partial\alpha}\ln\det\underline{M}_{i}^{\left(  \widehat
{n}\right)  }\left(  \varepsilon;\widehat{e}_{i}\right)  d\widehat{e}_{i}\;,
\end{equation}
with $\alpha=\vartheta$ or $\varphi$ and%
\begin{equation}
\underline{M}_{i}^{\left(  \widehat{n}\right)  }\left(  \varepsilon
;\widehat{e}_{i}\right)  =\underline{I}+\left(  \underline{t}_{i}\left(
\varepsilon;\widehat{e}_{i}\right)  ^{-1}-\underline{t}_{i,c}^{\left(
\widehat{n}\right)  }\left(  \varepsilon\right)  ^{-1}\right)  \underline
{\tau}_{ii,c}^{\left(  \widehat{n}\right)  }\left(  \varepsilon\right)  \;.
\end{equation}
By definition,%
\begin{align}
T_{\alpha}  &  =\frac{\operatorname{Im}}{\pi}\int d\varepsilon\,f_{FD}\left(
\varepsilon;\nu^{\left(  \widehat{n}\right)  }\right)  \sum_{i}\int
\lim_{\Delta\alpha\rightarrow0}\frac{P_{i}^{\left(  \widehat{n}+\Delta
\widehat{n}\right)  }\left(  \widehat{e}_{i}\right)  -P_{i}^{\left(
\widehat{n}\right)  }\left(  \widehat{e}_{i}\right)  }{\Delta\alpha}\ln
\det\underline{M}_{i}^{\left(  \widehat{n}\right)  }\left(  \varepsilon
;\widehat{e}_{i}\right)  d\widehat{e}_{i}\\
&  =\frac{\operatorname{Im}}{\pi}\int d\varepsilon\,f_{FD}\left(
\varepsilon;\nu^{\left(  \widehat{n}\right)  }\right)  \sum_{i}\lim
_{\Delta\alpha\rightarrow0}\frac{1}{\Delta\alpha}\left\{  \int P_{i}^{\left(
\widehat{n}+\Delta\widehat{n}\right)  }\left(  \widehat{e}_{i}\right)  \ln
\det\underline{M}_{i}^{\left(  \widehat{n}\right)  }\left(  \varepsilon
;\widehat{e}_{i}\right)  d\widehat{e}_{i}\right. \\
&
\begin{array}
[c]{cccccccc}
&  &  &  &  &  &  &
\end{array}
\;\left.  -\int P_{i}^{\left(  \widehat{n}\right)  }\left(  \widehat{e}%
_{i}\right)  \ln\det\underline{M}_{i}^{\left(  \widehat{n}\right)  }\left(
\varepsilon;\widehat{e}_{i}\right)  d\widehat{e}_{i}\right\}  \;.\nonumber
\end{align}
Approaching $T=0$,%
\begin{equation}
P_{i}^{\left(  \widehat{n}\right)  }\left(  \widehat{e}_{i}\right)
\rightarrow\delta\left(  \widehat{n}-\widehat{e}_{i}\right)  \;,\;\underline
{t}_{i}\left(  \varepsilon;\widehat{e}_{i}\right)  \rightarrow\underline
{t}_{i}\left(  \varepsilon;\widehat{n}\right)  \;,
\end{equation}%
\begin{equation}
P_{i}^{\left(  \widehat{n}+\Delta\widehat{n}\right)  }\left(  \widehat{e}%
_{i}\right)  \rightarrow\delta\left(  \widehat{n}+\Delta\widehat{n}%
-\widehat{e}_{i}\right)  \;,\;\underline{t}_{i}\left(  \varepsilon;\widehat
{e}_{i}\right)  \rightarrow\underline{t}_{i}\left(  \varepsilon;\widehat
{n}+\Delta\widehat{n}\right)  \;,
\end{equation}
while%
\begin{equation}
\ln\det\underline{M}_{i}^{\left(  \widehat{n}\right)  }\left(  \varepsilon
;\widehat{e}_{i}\right)  \rightarrow Tr\left(  \left[  \underline{t}%
_{i}\left(  \varepsilon;\widehat{e}_{i}\right)  ^{-1}-\underline{t}%
_{i}^{\left(  \widehat{n}\right)  }\left(  \varepsilon\right)  ^{-1}\right]
\underline{\tau}_{ii}^{\left(  \widehat{n}\right)  }\left(  \varepsilon
\right)  \right)  \;.
\end{equation}
Therefore,%
\begin{align}
T_{\alpha}  &  =\frac{\operatorname{Im}}{\pi}\int d\varepsilon\,f_{FD}\left(
\varepsilon;\nu^{\left(  \widehat{n}\right)  }\right)  \sum_{i}\lim
_{\Delta\alpha\rightarrow0}\frac{1}{\Delta\alpha}\times\\
&  Tr\left(  \left[  \underline{t}_{i}\left(  \varepsilon;\widehat{n}%
+\Delta\widehat{n}\right)  ^{-1}-\underline{t}_{i}^{\left(  \widehat
{n}\right)  }\left(  \varepsilon\right)  ^{-1}\right]  \underline{\tau}%
_{ii,c}^{\left(  \widehat{n}\right)  }\left(  \varepsilon\right)  \right)
-Tr\left(  \left[  \underline{t}_{i}\left(  \varepsilon;\widehat{n}\right)
^{-1}-\underline{t}_{i}^{\left(  \widehat{n}\right)  }\left(  \varepsilon
\right)  ^{-1}\right]  \underline{\tau}_{ii}^{\left(  \widehat{n}\right)
}\left(  \varepsilon\right)  \right) \\
&  =\frac{\operatorname{Im}}{\pi}\int d\varepsilon\,f_{FD}\left(
\varepsilon;\nu^{\left(  \widehat{n}\right)  }\right)  \sum_{i}\lim
_{\Delta\alpha\rightarrow0}\frac{1}{\Delta\alpha}Tr\left(  \left[
\underline{t}_{i}\left(  \varepsilon;\widehat{n}+\Delta\widehat{n}\right)
^{-1}-\underline{t}_{i}\left(  \varepsilon;\widehat{n}\right)  ^{-1}\right]
\underline{\tau}_{ii}^{\left(  \widehat{n}\right)  }\left(  \varepsilon
\right)  \right) \\
&  =\frac{\operatorname{Im}}{\pi}\int d\varepsilon\,f_{FD}\left(
\varepsilon;\nu^{\left(  \widehat{n}\right)  }\right)  \sum_{i}Tr\left(
\frac{\partial\underline{t}_{i}\left(  \varepsilon;\widehat{n}\right)  ^{-1}%
}{\partial\alpha}\underline{\tau}_{ii}^{\left(  \widehat{n}\right)  }\left(
\varepsilon\right)  \right)  \;,
\end{align}
which is equivalent to Eq.(\ref{T0}).


\begin{thebibliography}{99}
\bibitem{Jansen} H.J.F.Jansen, Phys. Rev. B {\bf 59}, 4699 (1999).
\bibitem{Kubler} J.Kubler, {\it  Theory of itinerant electron magnetism}, (Oxford: Clarendon 2000).
\bibitem{review} J.B.Staunton, Rep.Prog. Phys. {\bf 57}, 1289, (1994).
\bibitem{Razee+99} S.S.A.Razee {\it et al.}, Phys. Rev. Lett. {\bf 82}, 5369, (1999).
\bibitem{Shick} A.B.Shick {\it et al.}, Phys. Rev. B {\bf 56}, R14259, (1997).
\bibitem{Till} T.Burkert {\it et al.},Phys. Rev. B {\bf 69}, 104426, (2004).
\bibitem{Laz} B.Lazarovits {\it et al.},J. Phys.: Cond. Matt.{\bf 16}, S5833, (2004).
\bibitem{Qian} X.Qian and W.Hubner, Phys. Rev. B {\bf 64}, 092402, (2001).
\bibitem{Cabria} I.Cabria {\it et al.}, Phys. Rev. B {\bf 63}, 104424, (2001).
\bibitem{micromag} e.g. H.Kronmuller {\it et al.}, J.Mag.Magn.Mat. {\bf 175}, 177, (1997); M.E.Schabes, J.Mag.Magn.Mat. {\bf 95}, 249-288, (1991).
\bibitem{AMR} D.V.Baxter {\it et al.}, Phys.Rev. B {\bf 65}, 212407, (2002); K.Hamaya 
{\it et al.}, J.Appl.Phys. {\bf 94}, 7657-61, (2003); A.B.Shick {\it et al.}, Phys.Rev. B 
{\bf 73}, 024418, (2006).
\bibitem{Callen} H.B.Callen and E.Callen, J.Phys.Chem.Solids, {\bf 27}, 1271,
(1966); N.Akulov, Z. Phys. {\bf 100}, 197, (1936); C.Zener, Phys. Rev. 
B{\bf 96}, 1335, (1954).
\bibitem{MAEvsT} J.B.Staunton {\it et al.}, Phys. Rev. Lett. {\bf 93}, 257204, (2004).
\bibitem{Mryasov} O.Mryasov {\it et al.}, Europhys.Lett. {\bf 69}, 805, (2005); R.Skomski {\it et al.},
J. Appl. Phys. {\bf 99}, 08E916, (2006).
\bibitem{Wu} X.W.Wu {\it et al.}, Appl. Phys. Lett. {\bf 82}, 3475, (2003).
\bibitem{Thiele2002} J.-U.Thiele {\it et al.}, J. Appl. Phys. {\bf 91}, 6595, (2002).
\bibitem{Okamoto} S.Okamoto {\it et al.}, Phys. Rev. B{\bf 66}, 024413, (2002).
\bibitem{Moriya} {\it Electron Correlations and Magnetism
 in Narrow Band System}, edited by T.~Moriya (Springer, N.Y., 1981).
\bibitem{DLM} B.L.Gyorffy {\it et al.}, J. Phys. F: Met. Phys. {\bf 15}, 1337 (1985). 
\bibitem{JBS+BLG} J.B.Staunton and B.L.Gyorffy, Phys. Rev. Lett. {\bf 69}, 371 (1992).
\bibitem{Sun} S.Sun {\it et al.}, Science {\bf 287}, 1989, (2000).
\bibitem{OHandley} R.C.O'Handley, {\it Modern Magnetic Materials}, (Wiley, 2000).
\bibitem{TAR} A.Lyberatos and K.Y.Guslienko, J. Appl. Phys. 
{\bf 94}, 1119, (2003); H.Saga {\it et al.}, Jpn. J. Appl. Phys. 
Part 1 {\bf 38}, 1839, (1999); M.Alex {\it et al.},
IEEE Trans. Magn. {\bf 37}, 1244, (2001).
\bibitem{FePt-MAE} O.A.Ivanov {\it et al.}, Fiz. Met. Metalloved. {\bf 35}, 92, (1973.
\bibitem{Farrow} R.F.Farrow {\it et al.}, J. Appl. Phys. {\bf 79}, 5967,(1996)..
\bibitem{Razee+97} S.S.A.Razee {\it et al.}, Phys. Rev. B{\bf 56}, 8082 (1997).
\bibitem{more} S.Ostanin {\it et al.}, Phys. Rev. B69, 064425, (2004).
\bibitem{ICNDR} S.Ostanin {\it et al.}, J.Appl.Phys.{\bf 93}, 453, (2003);
J.B.Staunton {\it et al.},J. Phys.: Cond. Matt.{\bf 16}, S5623, (2004).
\bibitem{Wang-torque} X.Wang {\it et al.}, Phys. Rev. B {\bf 54}, 61-64, (1996).
\bibitem{Russian-spinwave} A.I. Akhiezer, V.G.Baryakhtar and S.V.Peletminskii, {\it Spin Waves
and Magnetic Excitations}, (Amsterdam: North Holland), (1968).
\bibitem{EB+BG} E.Bruno and B.Ginatempo, Phys. Rev. B{\bf 55}, 12946, (1997).
\bibitem{KKR} J.Korringa, Physica {\bf 13}, 392, (1947); W.Kohn and N.Rostoker, Phys.Rev. {\bf 94}, 1111, (1954).
\bibitem{KKR-CPA} G.M.Stocks {\it et al.}, Phys. Rev. Lett. {\bf 41}, 34, (1978).
\bibitem{scf-kkr-cpa} G.M.Stocks and H.Winter, Z.Phys.B {\bf 46}, 95, (1982); 
D.D.Johnson {\it et al.}, Phys. Rev. Lett. {\bf 56}, 2088, (1986).
\bibitem{fccFe-2002}  S.S.A. Razee {\it et al.}, Phys. Rev. Lett. {\bf 88}, 147201, (2002).
\bibitem{MFL-EPL} M.F.Ling {\it et al.}, Europhys.Lett. {\bf 25}, 631, (1994).
\bibitem{DLM-alloys} J.B.Staunton {\it et al.},J. Phys.: Cond. Matt. {\bf  9}, 1281-1300, (1997).
\bibitem{invar-Entel} V.Crisan {\it et al.}, Phys. Rev. B {\bf 66}, 014416, (2002). 
\bibitem{LSIC} M.Lueders {\it et al.}, Phys. Rev. B {\bf 71}, 205109, (2005).
\bibitem{Ian-Gd} I.Hughes {\it et al.}, in preparation.
\bibitem{Dederichs} K.Sato  {\it et al.}, J. Phys.: Cond. Matt. {\bf 16}, S5491, (2004).
\bibitem{sweden} A.M.N.Niklasson {\it et al.}, Phys. Rev. B {\bf 67}, 235105, (2003).
\bibitem{Rowlands} D.A.Rowlands {\it et al.},  Phys. Rev. B {\bf 67}, 115109, (2003).
\bibitem{NLCPA-totE} D.A.Rowlands {\it et al.},  Phys. Rev. B {\bf 73}, 165122, (2006).
\bibitem{Feynman} R.P.Feynman, Phys. Rev. {\bf 97}, 660, (1955).
\bibitem{cpa} P.Soven, Phys.Rev. {\bf 156}, 809, (1967).
\bibitem{Messiah} A.Messiah, {\it Quantum Mechanics}, (Amsterdam: North Holland), (1965).
\bibitem{Strange1984} P.Strange {\it et al.}, J.Phys. C {\bf 17}, 3355-71, (1984).
\bibitem{Gyorffy+Stott} B.L.Gyorffy and M.J.Stott, in {\it Band Structure Spectroscopy of 
Metals and Alloys}, eds.: D.J.Fabian and L.M.Watson, (Academic Press, New York), (1973).
\bibitem{Ginatempo+JBS} B.Ginatempo and J.B.Staunton, J.Phys.F {\bf 18}, 1827-37, (1988).
\bibitem{Lloyd} P.Lloyd and P.R.Best, J.Phys.C {\bf 8}, 3752, (1975).
\bibitem{Faulkner+Stocks} J.S.Faulkner and G.M.Stocks, Phys. Rev. B {\bf 21}, 3222, (1980).
\bibitem{Laszlo} A.Buruzs {\it et al.}, submitted to J.Mag.Magn.Mat. (2006).
\bibitem{dynsusc-bigpaper} J.B.Staunton {\it et al.}, Phys. Rev. B {\bf 62}, 1075-82, (2000).
\bibitem{TcFePd} L.Wang {\it et al.}, J.Appl.Phys. {\bf 95}, 7483-5, (2004).
\bibitem{FePd-expt} A.Ye.Yermakov {\it et al.}, Fiz. Met. Metall. {\bf 69}, Pt.5, 198, (1990);
H.Shima {\it et al.}, J. Mag. Magn. Mat. {\bf 272}, Part 3, 2173, (2004)
\bibitem{FePd-theory} G.H.O.Daalderop {\it et al.}, Phys. Rev. B {\bf 44}, 12054, (1991); I.V.Solovyev 
{\it et al.}, Phys. Rev. B {\bf 52}, 13419, (1995); I.Galanakis {\it et al.}, Phys. Rev. B {\bf 62}, 6475,
(2000); D.Garcia {\it et al.}, Phys. Rev. B {\bf 63}, 104421, (2001).
\bibitem{Burkert} T.Burkert {\it et al.}, Phys. Rev. B {\bf 71}, 134411, (2005).
\bibitem{Antoniak} C.Antoniak {\it et al.}, Europhys.Lett. {\bf 70}, 250-6, (2005).
\bibitem{SKKR} J.Zabloudil {\it et al.} in {\it Electron Scattering in Solid Matter}, Springer
Series in Solid State Sciences, {\bf 147} (Springer, Heidelberg, 2005).
\bibitem{Ebert} H.Ebert {\it et al.},Comp. Mat. Sci. {\bf 35}, 279-282, (2006).
\end{thebibliography}
\end{document}